\documentclass[conference,letterpaper]{IEEEtran}

\usepackage{booktabs}
\usepackage[ruled,linesnumbered]{algorithm2e}
\usepackage{amsmath}
\usepackage{amssymb}
\usepackage{multirow}
\usepackage[caption=false]{subfig}
\usepackage{color}
\usepackage{lipsum}
\usepackage[pdftex]{graphicx}
\usepackage[verbose]{cite}
\usepackage{url}
\usepackage{amsthm}

\theoremstyle{definition}
\newtheorem{definition}{Definition}[section]

\begin{document}

\title{Collusion-Secure Watermarking for Sequential Data}

\author{\IEEEauthorblockN{Arif Yilmaz}
\IEEEauthorblockA{Bilkent University\\
arif.yilmaz@bilkent.edu.tr}
\and
\IEEEauthorblockN{Erman Ayday}
\IEEEauthorblockA{Bilkent University\\
erman@cs.bilkent.edu.tr}
}


\maketitle

\begin{abstract}
In this work, we address the liability issues that may arise due to unauthorized sharing of personal data. We consider a scenario in which an individual shares his sequential data (such as genomic data or location patterns) with several service providers (SPs). In such a scenario, if his data is shared with other third parties without his consent, the individual wants to determine the service provider that is responsible for this unauthorized sharing. To provide this functionality, we propose a novel optimization-based watermarking scheme for sharing of sequential data. Thus, in the case of an unauthorized sharing of sensitive data, the proposed scheme can find the source of the leakage by checking the watermark inside the leaked data. In particular, the proposed schemes guarantees with a high probability that (i) the malicious SP that receives the data cannot understand the watermarked data points, (ii)  when more than one malicious SPs aggregate their data, they still cannot determine the watermarked data points, (iii) even if the unauthorized sharing involves only a portion of the original data or modified data (to damage the watermark), the corresponding malicious SP can be kept responsible for the leakage, and (iv) the added watermark is compliant with the nature of the corresponding data. That is, if there are inherent correlations in the data, the added watermark still preserves such correlations. Watermarking typically means changing certain parts of the data, and hence it may have negative effects on data utility. The proposed scheme also minimizes such utility loss  while it provides the aforementioned security guarantees. Furthermore, we conduct a case study of the proposed scheme on genomic data and show the security and utility guarantees of the proposed scheme.
\end{abstract}

\section{Introduction}

Sequential data includes time-series data such as location patterns, stock market data, speech, or ordered data such as genomic data. Individuals share different types of sequential data for several purposes, typically to receive personalized services from online service providers (SPs). For example, people share their continuous locations with map applications to use navigation services. Similarly, to provide location-based services, many online service providers motivate individuals to share their whereabouts. Recently, several direct-to-consumer SPs have emerged to collect individuals' genomic information to provide recreational services or to conduct research. The type of data collected and processed by these SPs may reveal significant privacy sensitive information about individuals. Location data of an individual may reveal information about his daily life such as his work and home addresses or his life style. Genomic data of an individual includes his personal and health-related data such as his physical characteristics and predisposition to diseases. Thus, the way these SPs handle the collected data poses a threat to individuals' privacy and it is crucial for individuals to have control on how their data is handled by the SPs.
	
As an individual shares his personal data with an SP for a particular purpose, he wants to make sure that his data will not be observed by other third parties. Privacy leakage occurs when personal data of individuals is further shared by an SP with other third parties (e.g., for financial benefit). To deter the SPs from such unauthorized sharing, it is required to develop technical solutions that would keep them liable for such unauthorized sharing (e.g., by connecting the unauthorized sharing to its source). One well-known tool for such scenarios is watermarking. An individual may add a unique watermark into his data before he shares it with each SP, and if his data is further shared without his authorization, he can associate the unauthorized sharing to the corresponding SP.
	
Watermarking is a well-known technique to address the liability issues for multimedia data \cite{buyer-seller}. Using the high amount of redundancy in the data and the fact that human eye cannot differentiate slight differences between the pixel values, watermark is inserted into multimedia data by changing some pixel values. However, watermarking is not a straightforward technique for sequential data such as location patterns or genomic data. To insert watermark into sequential data, some data points should be modified according to the watermark. In the case of location data, the individual may alter some of his actual location data as the watermark. In the case of genomic data, one may change the values of some nucleotides as the watermark. In both examples, original data is modified to add the watermark. Thus, watermarking sequential data while preserving data utility has unique challenges.
	
Another challenge for watermarking sequential data is the identifiability of the watermark. An individual cannot identify the SP that is responsible for the data leakage if the SP finds the watermark inserted data points locations and removes the watermarks before the unauthorized sharing. Thus, as opposed to multimedia data (in which watermark can be hidden in the redundancy in the data), it is more challenging to make sure that the SP cannot identify the watermarked data points in sequential data. An SP may utilize different types of auxiliary information in order to determine the watermark in the data. One type of such auxiliary information may be the inherent correlations in the data. Location patterns are correlated in both time and space. Similarly, genomic data carries inherent correlations (referred as linkage disequilibrium) inside. Thus, an SP can identify the watermarked data points by identifying the points that violate the inherent correlations in the data. Another type of auxiliary information is the data shared by the individual with other SPs. Multiple SPs may collect the same sequential data from the same individual (with different watermarks patterns) and they may compare their collected data in order to identify the watermark points with higher probability. Furthermore, even if the SP partially shares a portion of the data (rather than sharing the whole data of the individual) or modify the data (to damage the watermark), it should still be associated to this unauthorized sharing with a high probability.\\
	
\emph{Contributions.} To address these security and utility challenges, we propose a novel watermarking-based scheme to share sequential data. The adoption of watermarking prevents unauthorized sharing of sequential data by the SPs. In such a case, the data owner (or a third party) can associate the source of the leakage to the corresponding SP (or SPs). As discussed, watermarking is already commonly used to prevent illegal copies of multimedia data. The main contributions of the proposed work are summarized as follows:
\begin{itemize}
\item We propose a novel collusion-secure watermarking scheme for sequential data.
\item The proposed scheme minimizes the probability for the identifiability of the watermark by the SPs. We show that even when multiple SPs join their data together or they use the knowledge of inherent correlations in the data the watermark cannot be identified (with a high probability).
    \color{black}
\item We show that the SPs that are responsible for the unauthorized sharing can be detected with a high probability even when they share a portion of the data or when they modify the data in order to damage the watermark. We also show relationship between the probabilistic limits of this detection and the shared portion of data.\color{black}
\item While providing these security (or robustness) guarantees, the proposed system also minimizes the utility loss in the sequential data due to watermarking.
\item We also implement and evaluate the proposed scheme for genomic data sharing. The main motivations to choose genomic data sharing as the use case are as follows: (i) genomic data includes privacy-sensitive information such as predisposition to diseases~\cite{survey_naveed}, (ii) it is not revokable, and hence it is crucial to make sure that it is not leaked, and (iii) it has inherent correlations that makes watermarking even more challenging.
\end{itemize}
	
We believe that the proposed scheme will deter the SPs from unauthorized sharing of individual data with third parties. The rest of the paper is organized as follows. In the next section, we discuss the related work on watermarking and security and privacy of genomic data. In Section~\ref{sec:Problem Definition}, we introduce the data model, the system model, and the threat model. In~Section~\ref{sec:Proposed Solution}, we provide the details of the proposed solution. In Section~\ref{sec:Evaluation}, we evaluate the security of the proposed watermarking algorithm. In Section~\ref{sec:discussion}, we discuss potential extensions of the proposed scheme and possible future research directions. Finally, in Section~\ref{sec:Conclusion}, we conclude the paper.
	
\section{Related Work} \label{sec:Related Work}
	
\emph{Watermarking.} Digital watermarking is the act of hiding a message related to a digital signal (e.g., an image, song, or video) within the signal itself~\cite{cox2002digital}. It is closely related to steganography, both of them hide a message inside a digital signal. The difference between digital watermarking and the steganography is their goal. Watermarking hides a message related to the actual content of the digital signal, but in steganography the message and the actual content of the digital signal are not related, the digital signal is merely used as a cover for the message.
	
Digital watermarks can be used for copy protection and copy deterrence. Bloom et al. and Maes et al. proposed a system to protect copyrights on multimedia content in digital video disk (DVD)~\cite{Bloom1999,kalker}. A watermark is inserted into the content as a counter of copy number. Every time the content is copied, the watermark can be modified meaning that the counter is incremented. If the counter reaches a predefined limit, the hardware would not create further copies of the multimedia content. Memon and Wong proposed a system that uses watermark to detect illegal copy of digital content~\cite{buyer-seller}. This system guarantees that the seller of the digital content cannot share the digital content in an unauthorized way and it blames the buyer for the illegal copy.
	
Mainly due to the redundancy in multimedia content, watermarks are included in multimedia content relatively easier compared to informative text. Adding or removing a word or a character to informative text (as the watermark) can be easily detected by analysing the text. Several works proposed different systems to insert watermark into text documents~\cite{Brassil94hidinginformation,Brassil1995,Brassil1999}. Such systems typically include the watermark into the text in two ways: (i) the first technique is the line-shift algorithm which moves a line upward or downward (left or right) depending on watermark, and (ii) the second one is the word-shift algorithm which moves the words horizontally, thus expanding spaces to embed the watermark. Furthermore, Atallah et al. proposed a natural language watermarking scheme using the syntactic structure of the text~\cite{Atallah1,Atallah2}. Similarly, Topkara et al. proposed a sentence based text watermark algorithm that relies on multiple features of each sentence and exploits the notion of orthogonality between features~\cite{Topkara1}. These works have two weakness compared to our proposed scheme. They insert the watermark to the format of the text (e.g., by expanding spaces). If one removes the formatting, watermark disappears, and hence protection of the data is lost. Additionally, none of these works protect the watermark against collusion attacks. That is, if two or more different watermarked data are joined (by different receivers of the same data), watermark inserted data points can be easily identified.
	
Boneh and Shaw proposed a general fingerprint (watermark) solution that is secure against collusion~\cite{boneh1}. Their scheme constructs fingerprints in such a way that no coalition of attackers can find a fingerprint. However, there are still some practical drawbacks of this scheme. First, the length of the fingerprint is a problem. The scheme does not consider the utility loss of the data when it adds fingerprint into the data. To guarantee security against collusion, fingerprint length may be very long. Furthermore, this scheme does not consider the inherent correlations in the data. Sequential data may have correlations so that attackers may find the fingerprints by checking the correlated data points. As discussed, we address these drawbacks in our proposed scheme.\\
	
\emph{Security and privacy of genomic data.} Research on security and privacy of genomic data has gained significant pace over the last few years. Several attacks have been proposed showing vulnerability of genomic data. Notably, it has been shown that standard anonymization techniques are ineffective on genomic data~\cite{Gymrec_Science,Homer_2008}. Also, Humbert et al. evaluated the kin genomic privacy of an individual threatened by his relatives~\cite{humbert2013}.
	
As a response to these attacks, several protection mechanisms have also been proposed. Many researchers proposed cryptographic solutions to process genomic data in a privacy-preserving way~\cite{Pastoriza_CCS_2007,Blanton_dbsec_2010,Chen_NDSS_2012}. Baldi et al. and Ayday et al. proposed techniques for the privacy-preserving use of genomic data in clinical settings~\cite{Baldi_CCS_2011,ayday2013wpes}. Furthermore, Karvelas et al. proposed using the oblivious RAM mechanisms to access genomic data~\cite{Karvelas:2014:PWG:2665943.2665962}. Huang et al. proposed an information-theoretical technique for secure storage of genomic data~\cite{huang2015_genoguard}. Recently, Wang et al. proposed private edit distance protocols to find similar patients (across several hospitals)~\cite{Wang:2009:PGC:1653662.1653703}.
	
Using genomic data in a privacy-preserving way for research purposes has been also an important research topic. For this purpose, Johnson and Shmatikov proposed the use of differential privacy concept. Other works also proposed the use of homomorphic encryption and secure hardware for the same purpose~\cite{Kantarcioglu:2008:CAS:2222946.2223571,Canim_2012}. In this work, different from all previous work on genomic security and privacy, we propose a novel watermarking technique that addresses the liability issues on sequential data (including genomic data) in case of unauthorized sharing.

\section{Problem Definition} \label{sec:Problem Definition}
Here, we describe the data model, system model, and the threat model. Frequently used symbols and notations are presented in Table~\ref{table:notationTable}.
	\begin{table}[]
		\centering
		\begin{tabular}{|c|l|}
			\hline
			$x_1$, $\cdots$ , $x_\ell$ & Set of ordered data points \\ \hline
			$d_1,\cdots,d_m$ & Possible values (states) of a data point \\ \hline
			$I_i$ & \begin{tabular}[c]{@{}l@{}}Index set of the data points that are \\ shared with the SP $i$\end{tabular} \\ \hline
			$D_{I_i}$ & Set of data points in $I_i$ \\ \hline
			$W_{I_i}$ & Set of data points in $I_i$ after watermarking\\ \hline
            $Z_{I_i}$ & Set of watermarked data points in $W_{I_i}$\\ \hline
		\end{tabular}
		\caption{Frequently used symbols and notations.}
		\label{table:notationTable}
	\end{table}

\subsection{Data Model} \label{sec:The Data Model}
Sequential data consists of ordered data points $x_1$, ..., $x_\ell$, where $\ell$ is the length of the data. The value of a data point $x_i$ can be in different states from the set $\{d_1,\cdots,d_m\}$ according to the type of the data. For instance, $x_i$ can be coordinate pairs in terms of latitude and longitude for location data, it can be location semantics (e.g., cafe or restaurant) for check-in data, or it can be the value of a nucleotide or point mutation for genomic data.
	
We approach the problem for two general sequential data types: (i) sequential data with no correlations in which data points are independent and identically distributed. In this type, value of a data point cannot be predicted using the values of other data points. Sparse check-in data might be a good example for this type. And, (ii) sequential data with correlations between the data points. This correlation between data points may vary based on the type of data. For example, consecutive data points that are collected with small differences in time may be correlated in location patterns. That is, an individual's location at time $t$ can be estimated if his locations at time $(t-1)$ and/or $(t+1)$ are known. In genomic data, point mutations (e.g., single nucleotide polymorphisms or SNPs\footnote{We provide a brief background on genomics in Section~\ref{sec:Evaluation}.}) may have pairwise correlations between each other. Such pairwise correlations are referred as linkage disequilibrium \cite{Slatkin2008} and they are not necessarily between consecutive data points. The correlation value may differ based on the state of each data point and correlation between the data points is typically asymmetric. Furthermore, it has been shown that  correlations in human genome can also be of higher order~\cite{samani2015}. For the clarity of the presentation, we first build our solution for uncorrelated sequential data and then extend it for correlated data.
\subsection{System Model} \label{sec:The System Model}
	
\begin{figure}[h!]
\centering
		\includegraphics[width=0.5\textwidth]{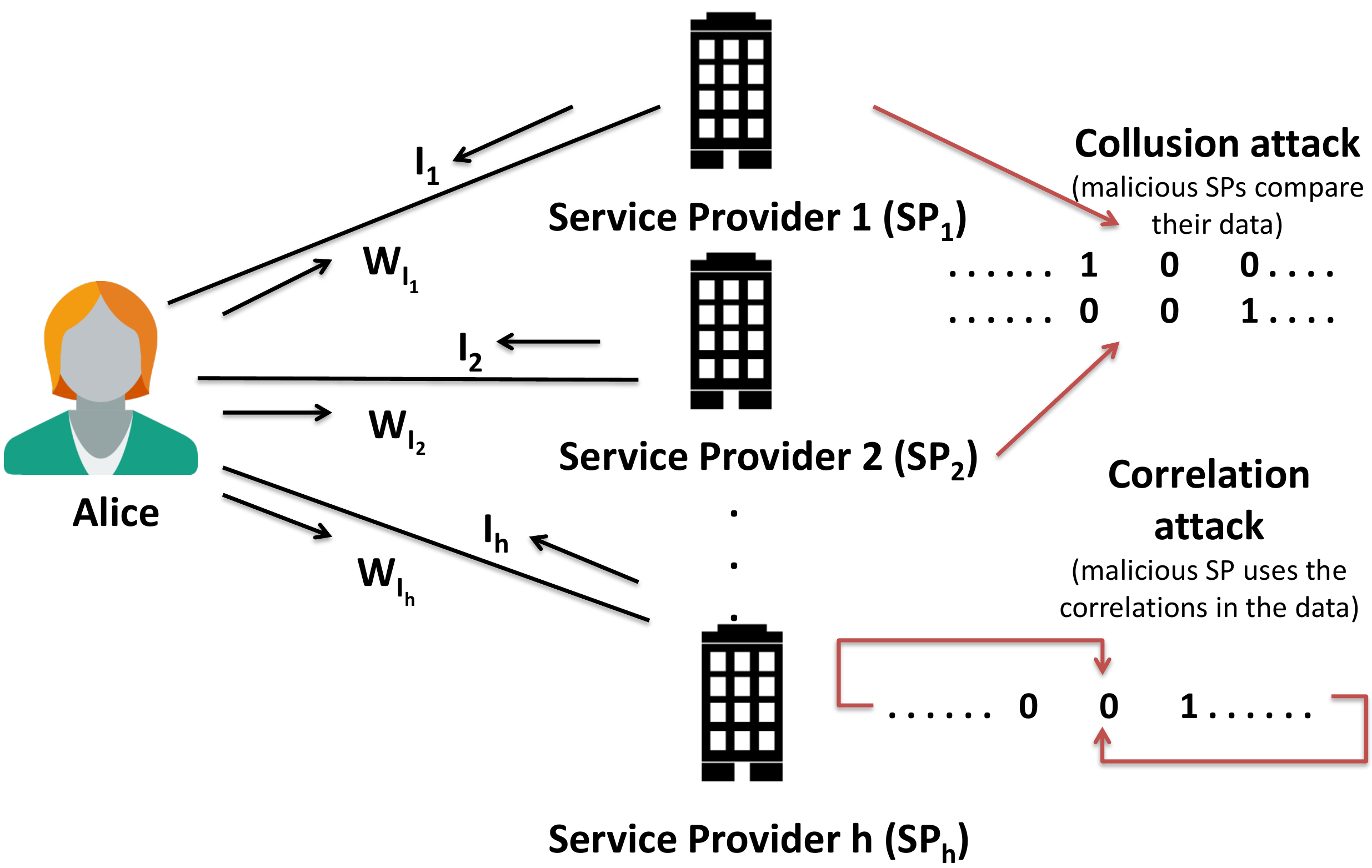}
		\caption{Overview of the system and threat models.}
		\label{fig:system_model}
\end{figure}

We consider a system between a data owner (Alice) and multiple service providers (SPs) as shown in Figure~\ref{fig:system_model}. For genomic data, the SP can be a medical institution, a genetic researcher, or direct-to-customer service provider. For  location data, the SP can be any location-based service provider. \color{black} In the description of the scheme, for clarity, we give illustrative examples on binary data but the proposed scheme can be extended for non-binary data. In fact, for the evaluation of the proposed scheme, we focus on the point mutations in genomic data that may have values from $\{0,1,2\}$. \color{black} Alice shares parts of her data with the SPs to receive different types of services. \color{black} Note that the part Alice shares with each SP may be different and we do not need same data to be shared with each SP. When we talk about the collusion attack (as will be detailed in the next section), we consider the intersection of the data parts owned by all malicious SPs.\color{black}

On one hand, when Alice shares her data with an SP, she wants to make sure that her data will not be shared with other third parties by the corresponding SP. In the case of further unauthorized sharing, she wants to know the SP that is responsible from this leak. Therefore, whenever Alice shares her data with a different SP, she inserts a unique watermark into it. On the other hand, an SP may share Alice's data with third parties without the consent of Alice. While doing so, to avoid being detected, the SP wants to detect and remove the watermark from the data. Instead of sharing the whole data with a third party, an SP may also share a certain portion of Alice's data to reduce the risk of detection (but compromising from the shared data amount). \color{black} Similarly, malicious SP (or SPs) may try to damage the watermark by modifying the data. \color{black} Furthermore, two or more SPs may join their data to detect the watermarked points. Security of the watermarking scheme increases (against the attacks discussed in the next section) as the length of the watermark increases. However, a long watermark causes significant modification on the original data, and hence decreases the utility of the shared data. In our proposed scheme, utility loss in the data is minimized while the watermarking scheme is still robust against the potential attacks with high probability.
	
\subsection{Threat Model} \label{sec:The Threat Model}
	
Here, we discuss the attacks we consider against the proposed watermarking scheme and our definitions for watermark robustness under these attacks.

\subsubsection{Attack models}

We consider the following attacks against the proposed watermarking scheme:\footnote{We assume there is secure communication between Alice and the SPs. Therefore, an outsider attacker can neither eavesdrop nor modify the data.}
	
\noindent\textbf{Single SP attack on uncorrelated data:} If the SP wants to leak Alice's data, the SP should find and remove the watermarks from the data so that Alice cannot blame the SP for this leak. Assume that Alice shares her (uncorrelated) sequential data of length $\ell$ with the SP and she includes a watermark of length $w$ into this data. Since data is uncorrelated, each data point is independent from other, and hence for each data point, the probability of being watermarked  is $w/\ell$. We also assume that the SP does not know any auxiliary information about the data owner. Therefore, it cannot find the watermark inserted data points with a higher probability. \color{black} Alternatively, instead of trying to detect the watermark, the malicious SP may also modify the data in order to damage the watermark.\color{black}
	
\noindent\textbf{Correlation attack:} If an SP has correlated data points and it also knows the corresponding correlation values, is may identify the watermarked points with higher probability. To be general, we assume pairwise, asymmetric correlations between different states of data points. The proposed scheme can be extended to other scenarios (e.g., higher order correlations or symmetric correlations) similarly.

As discussed, a data point may take values from the set $\{d_1, d_2, ...,d_m\}$. If $d_\alpha$ state of $x_i$ (i.e., $x_i = d_\alpha$)  is correlated with $d_\beta$ state of $x_j$ (i.e., $x_j = d_\beta$), then $Pr(x_i=d_\alpha | x_j=d_\beta)$ is high, but the opposite does not need to hold (i.e., $Pr(x_j=d_\beta | x_i=d_\alpha)$ does not need to be high). Note that $d_\alpha$ state $x_i$ may be in pairwise correlation with other data points as well. We consider all possible pairwise correlations between different states of all data points in our analysis.
	
Following the above example, assume the SP has one of the correlated data points as $x_j = d_\beta$, but $x_i = d_\gamma$ (where $d_\gamma \neq d_\alpha$). Then, the SP can conclude that $x_i$ is watermarked with probability $p(x_i^w)=Pr(x_i = d_\alpha | x_j = d_\beta)$. If $d_\alpha$ state of $x_i$ is also correlated with other data points (that the SP can observe), then the SP computes the watermark probability on $x_i$ as the maximum of these probabilities. Similarly, $d_\gamma$ state of $x_i$ may also be correlated with other data points that the SP can observe. Since, $x_i = d_\gamma$, such correlations imply that data point $x_i$ is not watermarked. Using such correlations, the SP also computes the probability that $x_i$ is not watermarked, $p(x_i^n)$. Eventually, the SP computes the probability of data point $x_i$ being watermarked as $(p(x_i^w)-p(x_i^n))$ (if the computed value is negative, we make it zero). We further explain this correlation model in Section~\ref{sec:For Data With Correlations}.
	
Once the SP determines the probability of being watermarked for each data point, it sorts them based on the computed probabilities, and identifies the watermarked data points as the ones with the highest probabilities. We assume that the SP knows the watermarking algorithm, and hence the length of the watermark ($w$). Thus, the SP may chose $w$ data points corresponding to the $w$ highest probabilities to infer the watermarked data points in the shared data. Note that there may be less that $w$ data points with positive probabilities. In such cases, the malicious SP (or SPs) infer the remaining watermarked points using either the single SP or collusion attack.
	
\noindent\textbf{Collusion attack:} Multiple SPs that receive the same data portion (from the same data owner) with different watermark patterns may join their data to identify the watermarked points with higher probability. In such a scenario, when the SPs vertically align their data points, they will observe some data points with different states. Such data points will definitely be marked as watermarked data points by the SPs (with different probabilities, as will be discussed later), and hence they will have more chance to identify the watermarked positions. Note that collusion attack may also benefit from the correlation attack and each SP may first run the correlation attack on their data before they join the data for the collusion attack.  We also evaluate the security of the proposed scheme against such an attack. \color{black} Similar to the single SP attack, malicious SPs may also try to modify the data in order to damage the watermark.\color{black}
	
\color{black}

\subsubsection{Watermark robustness}\label{sec:robustness}

``Robustness'' and ``security'' terms have been used interchangeably for watermarking schemes in different works. Adelsbach et al. provide formal definitions for watermark robustness~\cite{Adelsbach:2006:CMW:1759048.1759060}. Different from our work, in~\cite{Adelsbach:2006:CMW:1759048.1759060}, authors consider watermarking mechanisms that use a secret embedding key (that is used when adding watermark to the data). Thus, Adelsbach et al. mainly consider computational robustness that relies on the computational hardness of a problem. They define watermark robustness as the information of the watermark that is revealed to the adversary and watermark security as the information revealed about the secret embedding key.

They consider two adversary models (passive and active) and define watermark robustness for both. Robustness for passive adversary requires watermark to remain detectable when data is maliciously modified as long as the watermarked data is perceptibly similar to the original data. This similarity metric is defined differently for each different application and we use the data utility value to measure the difference (similarity) between the watermarked and original data. This definition is similar to our robustness requirement, however in~\cite{Adelsbach:2006:CMW:1759048.1759060}, the authors do not consider collusion and correlation attacks for the watermark robustness. Robustness for active adversary, on the other hand, considers an adversary having access to embedder and detector including the corresponding keys. Inspired from~\cite{Adelsbach:2006:CMW:1759048.1759060}, we come up with the following robustness definitions for the proposed watermarking scheme.

\noindent\textbf{Robustness against watermark inference:} This property states that watermark should not be inferred by the malicious SP (or SPs) via the aforementioned attack models. In the proposed scheme, inferring the watermark does not rely on a computationally hard problem; malicious SP (or SPs) probabilistically infer the watermark. Thus, we evaluate the proposed scheme for this property in terms of malicious SP's (or SP') inference probability for the added watermark. We provide the following definition to evaluate the robustness of a watermarking scheme against watermark inference.
\theoremstyle{definition}
\begin{definition}{\textbf{$p$-robustness against $f$-watermark inference.}}
A watermarking scheme is $p$-robust against $f$-watermark inference if probability of inferring at least $f$ fraction of the watermark ($0\leq{f}\leq1$, where $f=1$ means the whole watermark pattern) is smaller than $p$.
\end{definition}

\noindent\textbf{Robustness against watermark modification:} This property states that the malicious SP (or SPs) should not be able to modify the watermark in such a way that the watermark detection algorithm of the data owner misclassifies the source of the unauthorized data leakage. We evaluate the proposed scheme for this attribute in terms of precision and recall of the data owner to detect the malicious SP (or SPs) that leak her data. For this, we define ``false positive'' as watermark detection algorithm of the data owner classifying a non-malicious SP as a malicious one and ``false negative'' as watermark detection algorithm of the data owner classifying a malicious SP as a non-malicious one. We provide the following definition to evaluate the robustness of a watermarking scheme against watermark modification.
\theoremstyle{definition}
\begin{definition}{\textbf{$\rho/\epsilon$-robustness against watermark modification.}}
A watermarking scheme is $\rho/\epsilon$-robust against watermark modification if malicious SP (or SPs), by modifying the watermark, cannot decrease the precision and recall of the watermark detection algorithm below $\rho$ and $\epsilon$, respectively.
\end{definition}

For all the aforementioned attack models, we evaluate the proposed watermarking scheme based on its robustness. In Section~\ref{sec:Evaluation}, we show the limits of the proposed scheme for these definitions considering different variables.
 	
\color{black} 
\section{Proposed Solution} \label{sec:Proposed Solution}
	
Here, first we present an overview of the proposed protocol and then describe the details of the proposed watermarking algorithm.

\subsection{Protocol Overview}

\begin{figure}[h]
\centering
		\includegraphics[width=0.5\textwidth]{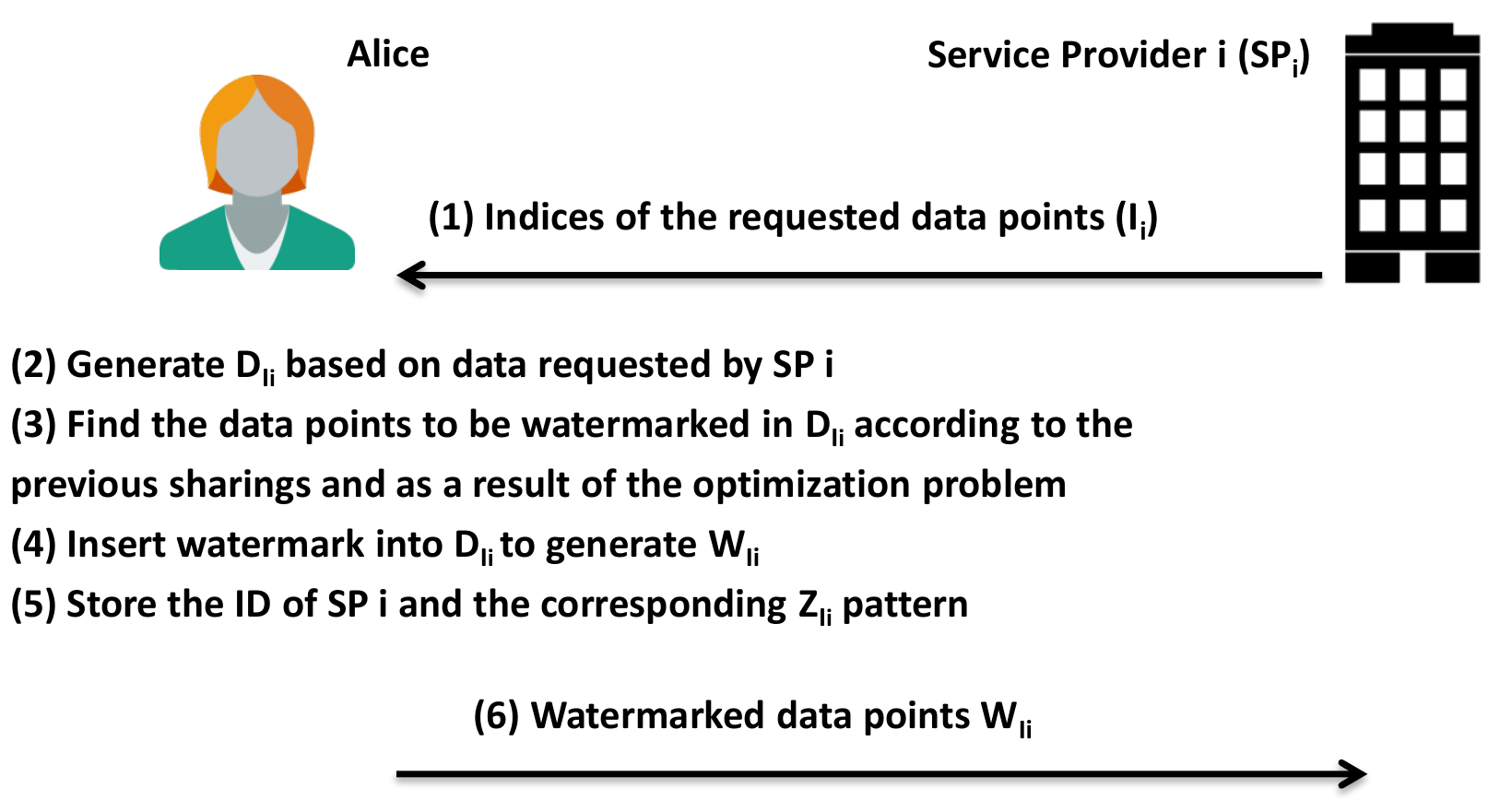}
		\caption{Data sharing protocol between Alice and a service provider}
        \label{fig:algorithm}
\end{figure}

When Alice wants to share her data with an SP $i$, they engage in the following protocol. The highlevel steps of the algorithm are also shown in Figure~\ref{fig:algorithm}.
	
\noindent(1) The SP $i$ sends the indices of Alice's data it requests, denoted by $I_i$.

\noindent(2) Alice generates $D_{I_i}=\bigcup_{i \in I_i} x_i$.

\noindent(3) Alice finds the data points to be watermarked considering her previous sharings of her data. This part is done using our proposed watermarking algorithm as described in detail in the next section.

\noindent(4) Alice inserts watermark into the data points in $D_{I_i}$ and generates the watermarked data $W_{I_i}$.

\noindent(5) Alice stores the ID of the SP and $Z_{I_i}$ (watermark pattern for the SP $i$).

\noindent(6) Alice sends $W_{I_i}$ to SP $i$.

\subsection{Watermarking Algorithm} \label{sec:The Watermark Insertion Algorithm}
	
In this section, we provide the details of our proposed watermarking algorithm. In particular, we describe the selection of data points to be watermarked in the sequential data so that the watermark will be secure against the attacks discussed in Section~\ref{sec:The Threat Model}.
	
We insert watermark into a data point by changing this data point's state. For instance, if data is binary, this change is from 0 to 1, or vice versa. If each data point can have states from the set $\{d_1,\cdots,d_m\}$, the change is from the current state to some other predefined state $d_j$. For the simplicity of discussion, we assume that for each data point, the watermarked state is predetermined. That is, whenever we decide to watermark a data point $x_i$, it is always changed to a predetermined state $d_j$.  This assumption can easily be extended to support changes into various states. In the following, we first detail our solution for sequential data that has no correlations (data points are independent from each other) and then, we will describe how to extend this for correlated sequential data.
	
\subsubsection{Sequential data without correlations} \label{sec:For Data With No Correlations}
	
Before giving the details of the proposed algorithm, we first provide the following notations that will facilitate the discussion.
\begin{itemize}
\item $n_i^h$: number of data points that are watermarked $i$ times when the whole data is shared with $h$ SPs.
\item $\hat{y}_i^h$: number of data points that are watermarked $i$ times when the whole data is shared $h$ times and will not be watermarked in the $(h+1)$-th sharing.
\item $y_i^h$: number of data points that are watermarked $i$ times when the whole data is shared $h$ times and will be watermarked in the $(h+1)$-th sharing.
\end{itemize}
	
\begin{figure}[htp]
\centering
	\includegraphics[clip,width=\columnwidth]{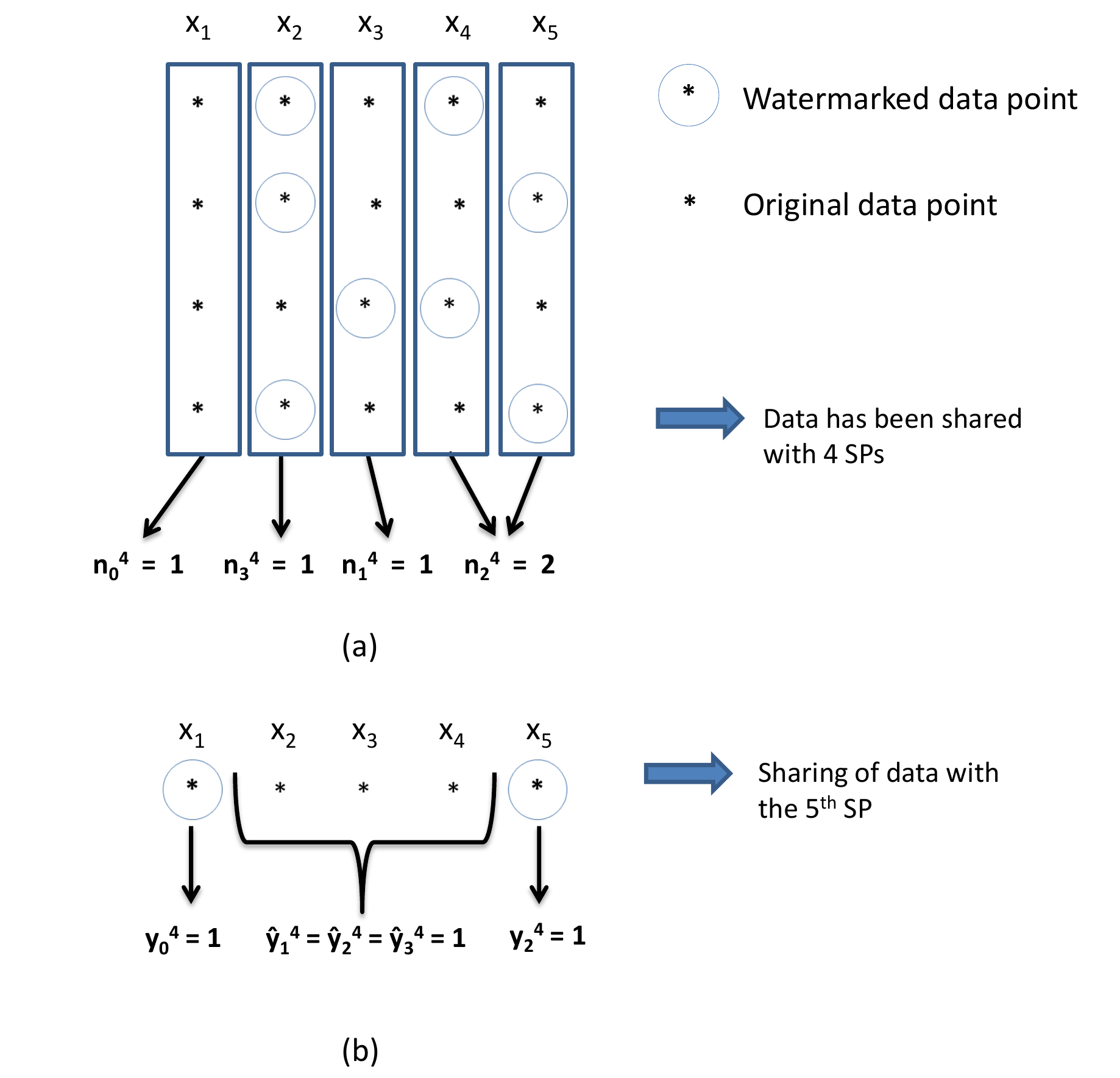}%
	\caption{Toy example for the notations in the watermark insertion algorithm.}
	\label{fig:toy_example_notation}
\end{figure}

We also provide a toy example in Figure~\ref{fig:toy_example_notation} to graphically represent these notations. In the toy example, Alice has a sequential data of length $5$ and she has already shared her data with $h=4$ SPs. The example also shows the instance when Alice shares her data with the $(h+1)$-th SP. In a nutshell, Alice, when she shares her data with the $(h+1)$-th SP, runs the watermarking algorithm to compute the $n_i^{h+1}$ values that would minimize the probability of the attacks discussed in Section~\ref{sec:The Threat Model}. Based on these values, she determines the data points to add the watermark.

The proposed algorithm is an iterative one. When Alice shares her data with a new SP, watermark locations in the data are determined for the new request according to the watermark patterns in previously shared data and then, the data points in the corresponding locations are modified. As discussed in Section~\ref{sec:The Threat Model}, a malicious SP may try to find the watermark inserted data points. A malicious SP needs auxiliary information about the data to increase its probability to find the watermark inserted data points. This information can be obtained from other SPs that received the same data from Alice with different watermark patterns. An example of the collusion attack may be described as follows.
	
\begin{figure}[h!]
\centering
		\includegraphics[width=0.5\textwidth]{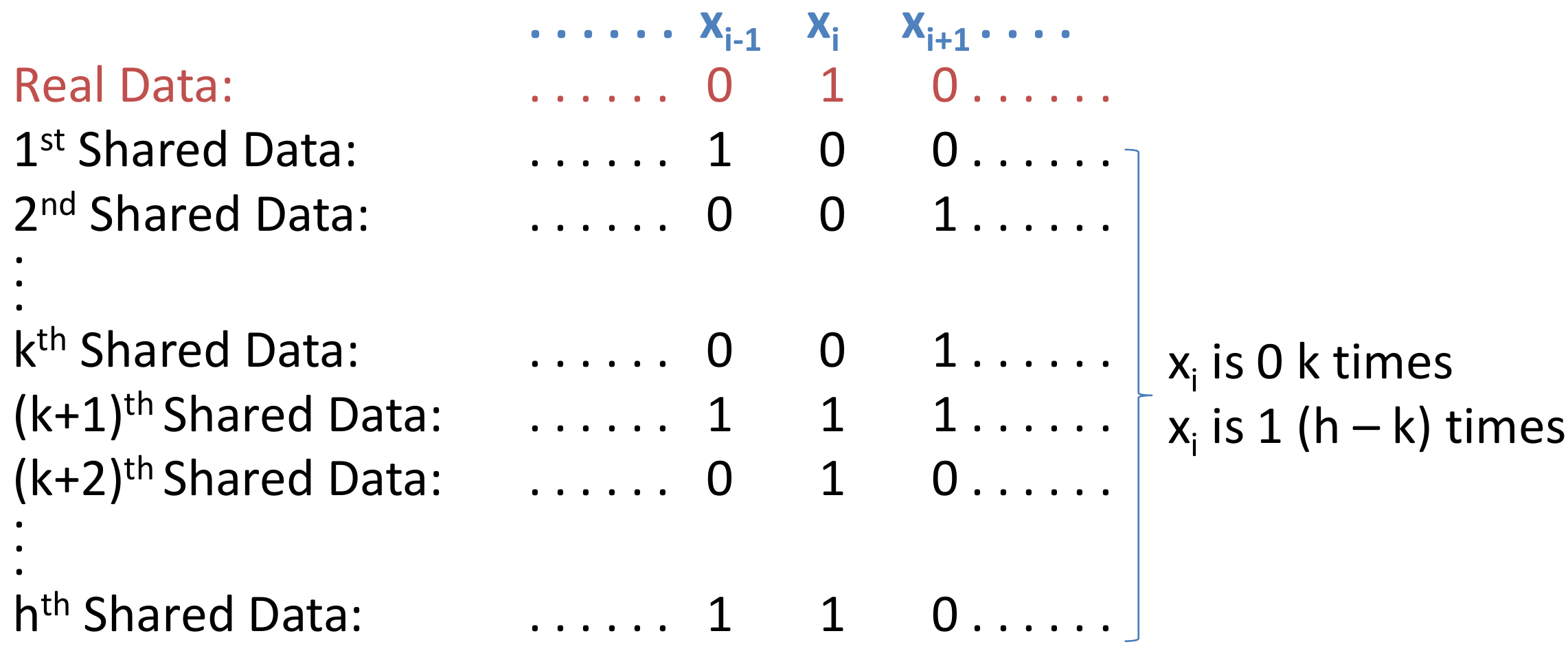}
		\caption{Collusion attack in which $h$ malicious SPs compare their data (belonging to the same individual). $i$-th data point ($x_i$) is 0 $k$ times and 1 $(h - k)$ times.}
		\label{fig:collusion_attack_1}
\end{figure}

For simplicity, assume that each data point can be either 0 or 1 and $h$ malicious SPs have the same data portion (belonging to Alice) with different watermark patterns as shown in Figure~\ref{fig:collusion_attack_1}. They vertically align their data portions, compare their data, and find the differences. For a data point $x_i$, they observe $k$ 0s and $(h-k)$ 1s (where $0\leq{k}\leq{h}$) and they conclude that the corresponding data point has either $k$ or $h-k$ watermarks.
	
We assume that the proposed watermarking algorithm is also known by the malicious SPs. Therefore, these $h$ SPs may run our proposed algorithm (as discussed next) and find $n_k^h$ and $n_{h-k}^h$ values. Once they have these values, they may compute that (i) the corresponding data point has $k$ watermarks with probability $\frac{n_k^h}{n_k^h + n_{h-k}^h}$, and (ii) $(h-k)$ watermarks with probability $\frac{n_{h-k}^h}{n_k^h + n_{h-k}^h}$. To their advantage, malicious SPs start inferring the watermark positions with higher probabilities during the attack.

Multiplication of such probabilities for every data point gives the probability of identifying the whole watermark inserted data points. To consider the worst-case scenario, we assume that malicious SPs have access to all previously shared data by Alice (i.e., all SPs that receive Alice's data collude) and all malicious SPs have the same data portion belonging to Alice (with different watermark patterns).\footnote{If malicious SPs have different data portions, they use the intersection of these portions for the collusion attack.} In our algorithm, watermarks are inserted into the watermark locations that minimizes the probability of identifying the whole watermarked points in the data. To do so, we solve a non-linear optimization problem to determine the data points to be watermarked at each data sharing instance of Alice. This problem can be formalized for the $(h+1)$-th sharing as follows:
\begin{equation*} \label{eqn:optimization_problem}
	\begin{split}
	& min \prod_{i=0}^{h+1} (\frac{n_i^{h+1}}{n_i^{h+1} + n_{h-i+1}^{h+1}})^{n_i^{h+1}}\\
	& s.t\\
	&(i) \quad \sum_{i=0}^{h+1} y_i^h = \mathrm{watermark}\: \mathrm{length}~ (w)\\
	&(ii) \quad n_0^{h+1} = \hat{y}_0^h\\
	&(iii) \quad n_{h+1}^{h+1} = y_h^h\\
	&(iv) \quad n_i^{h+1} = y_{i-1}^h + \hat{y}_i^h \:\mathrm{for}\: i = 1,...,h \\
	&(v) \quad \hat{y}_i^h + y_i^h = n_i^h\\
	&(vi) \quad y_i^h, \: \hat{y}_i^h \geqslant 0\:\mathrm{for}\: i = 0,...,h\\
    &(vii) \quad y_0^h > 0\\
	\end{split}
\end{equation*}

\begin{figure}[h!]
\centering
		\includegraphics[width=0.5\textwidth]{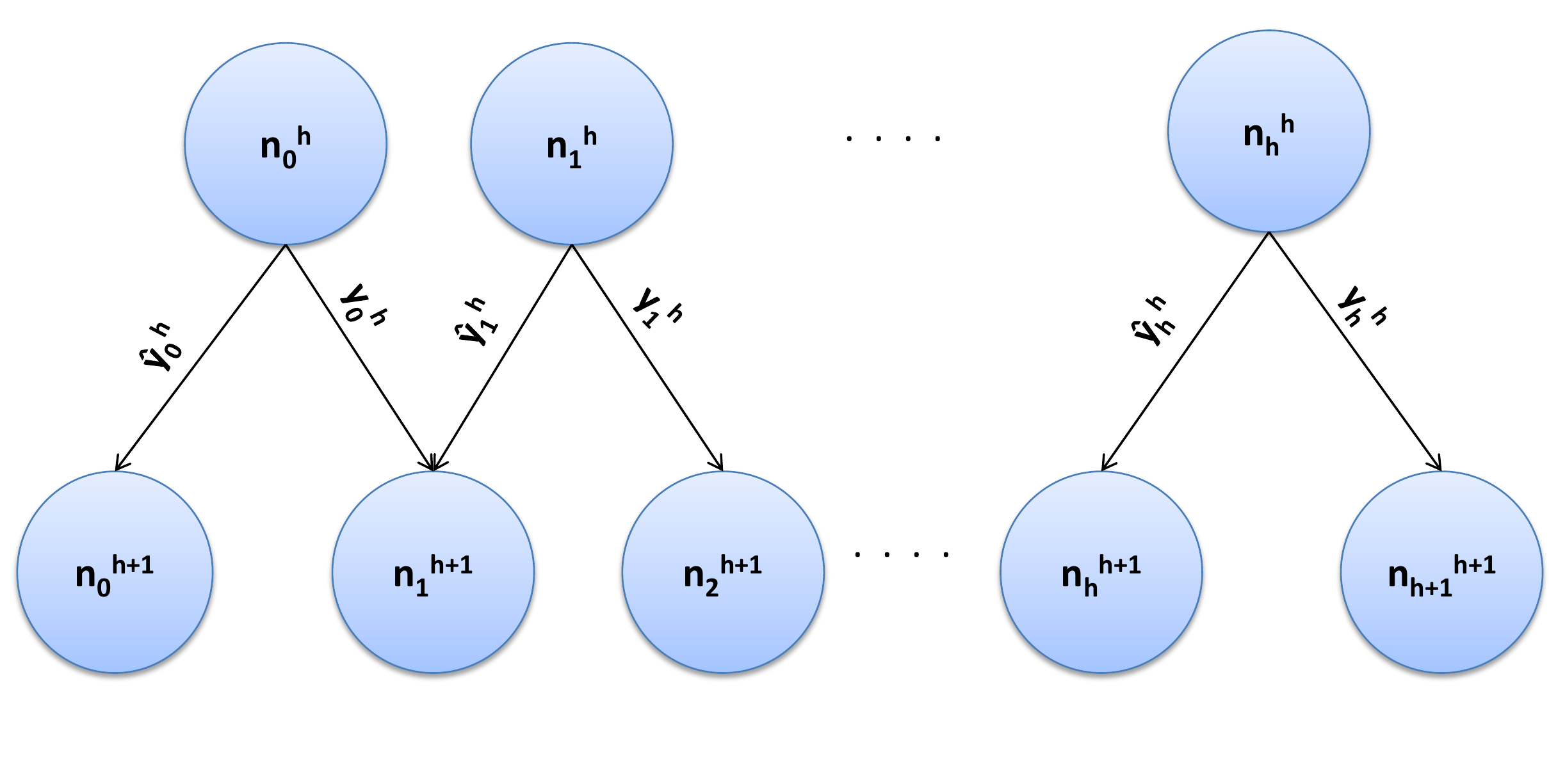}
		\caption{Relationship between $n_i^h$, $n_i^{h+1}$, $y_i^h$, and $\hat{y}_i^h$ values in the watermark insertion scheme.}
		\label{fig:wi_algorithm}
\end{figure}

Here, constraint $(i)$ determines the number of data points that we watermark. That is, the algorithm does not modify more data points than the limit defined in this constraint. Thus, for the tradeoff between the security of the watermark and data utility, the most important parts of the optimization problem are the objective function and constraint $(i)$. Constraints $(ii)$, $(iii)$, $(iv)$, and $(v)$ denote the relationship between $n_i^h$, $n_i^{h+1}$, $y_i^h$, and $\hat{y}_i^h$. In Figure~\ref{fig:wi_algorithm}, we show this relationship. Constraint $(vi)$ is used to prevent negative $y_i^h$ and $\hat{y}_i^h$ values. Finally, constraint $(vii)$ is to make sure that each SP has a unique watermark pattern. As the solution of this optimization problem, we obtain the $y_i^h$ and $\hat{y}_i^h$ values. Since in this scenario the data points are uncorrelated, we may choose any of the $n_i^h$ data points to insert the watermark. Note that $y_i^h$ $\leq$ $n_i^h$, and thus we will always have enough number of $i$-times watermark inserted data points among the previous ($h$) sharings of the data to insert watermark for the current ($(h+1)$-th) sharing.

\begin{figure}[h!]
\centering
		\includegraphics[width=0.5\textwidth]{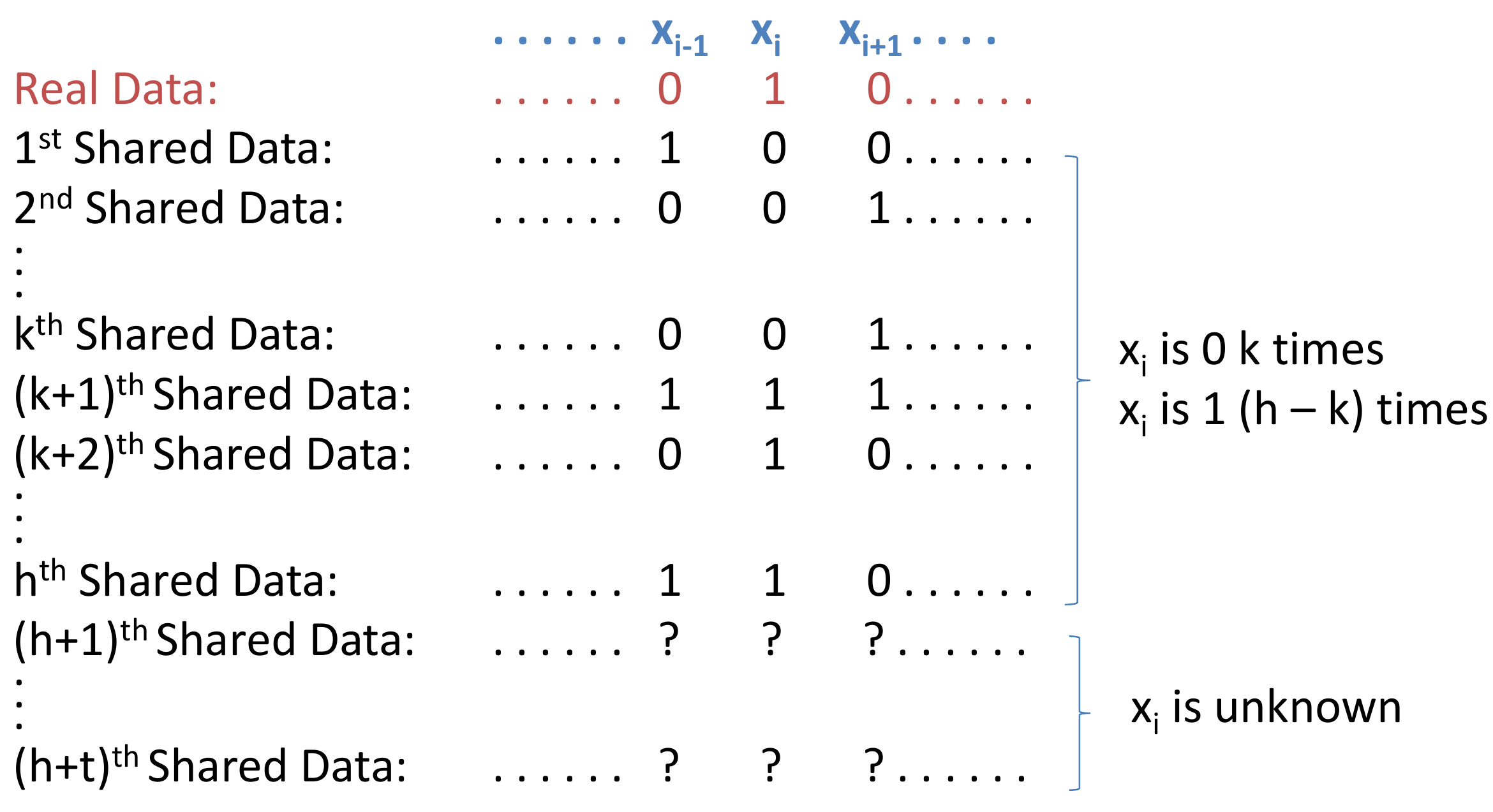}
		\caption{Collusion attack in which $h$ malicious SPs compare their data (belonging to same individual). Malicious SPs do not know the value of $x_i$ for $t$ sharings of the data with non-malicious SPs. }
		\label{fig:collusion_attack_2}
\end{figure}

\emph{Malicious SPs with partial knowledge.} In the above example, to illustrate the worst case scenario, we assume that $h$ malicious SPs correctly know that Alice shared her data totally $h$ times. However, this assumption may be too strong in practice. In practice, if $h$ malicious SPs join their data (belonging to the same individual), they just know that the data has been previously shared for at least $h$ times. To run their collusion attack, they should make an assumption about the total number of times Alice shared the same data before. For instance, if they assume that data has been previously shared (by Alice) for $h+t$ times, there will be $t$ unknown data points for each data position as shown in Figure~\ref{fig:collusion_attack_2}. Assuming data points take binary states, for each data location, the unknown $t$ data points contain $u$ 0s and $(t - u)$ 1s, where $0 \leq u \leq t$.
	
In this scenario, for a data location $x_i$ that has $k$ observed 0s and $(h-k)$ observed 1s, the colluding SPs follow the below steps to identify the watermark.
\begin{itemize}
\item There are $(k+u)$ 0s and $(h+t-k-u)$ 1s. That is, the corresponding data point is watermarked for either $(k + u)$ or $(h + t - k - u)$ times ($0 \leq u \leq t$).	
\item Colluding SPs run the algorithm in Section~\ref{sec:The Watermark Insertion Algorithm} and find $n_{k+u}^{h+t}$ and $n_{h+t-k-u}^{h+t}$ values.
\item The unknown $t$ data points contain $u$ 0s and $(t - u)$ 1s with the following probability:
	\begin{equation*}
		P_u = \frac{n_{k+u}^{h+t} + n_{h+t-k-u}^{h+t}}{\sum_{j=0}^{t}n_{k+j}^{h+t} + n_{h+t-k-j}^{h+t}}.
	\end{equation*}
\item Given the unknown $t$ data points contain $u$ 0s and $(t - u)$ 1s, $(h+t)$ data points have $(k + u)$ watermarks with probability
	\begin{equation*}
		p_{k+u} = \frac{n_{k+u}^{h+t}}{n_{k+u}^{h+t} + n_{h+t-k-u}^{h+t}}.
	\end{equation*}
\item Similarly, given the unknown $t$ data points contain $u$ 0s and $(t - u)$ 1s, $(h+t)$ data points have $(h + t - k - u)$ watermarks with probability
	\begin{equation*}
		p_{h+t-k-u} = \frac{n_{h+t-k-u}^{h+t}}{n_{k+u}^{h+t} + n_{h+t-k-u}^{h+t}}.
	\end{equation*}
\item Finally, malicious SPs conclude that these $(h+t)$ data points have $(k + u)$ watermarks with probability $P_u \cdot p_{k+u}$
		and these data points have $(h + t - k - u)$ watermarks with probability $P_u \cdot p_{h+t-k-u}$.
	\end{itemize}
Using these computed probabilities, the colluding SPs can probabilistically identify the watermark for each data location. Note that our proposed watermarking algorithm also minimizes the probability of this attack. We study this scenario and show the security of our proposed scheme against this attack in Section~\ref{sec:Evaluation}.

\color{black}
	
\subsubsection{Sequential data with correlations} \label{sec:For Data With Correlations}

In Section~\ref{sec:For Data With No Correlations}, malicious SPs do not have any auxiliary information about the data. Sequential data generally consists of data points that are correlated. As discussed in Section~\ref{sec:The Threat Model}, if the sequential data has correlations inside, attackers may find the watermark inserted data points easier (with higher probability). Therefore, while adding watermark to (correlated) sequential data, we should also make sure that strong correlations in the data would not be disturbed.
	
We insert watermarks into data in such a way that the correlations inside the sequential data are preserved. Similar to uncorrelated data, we follow the protocol and solve the optimization problem given in Section~\ref{sec:For Data With No Correlations}. In this scenario, the main difference is the way we choose the data points to insert watermark.
	
By solving the optimization problem in Section~\ref{sec:For Data With No Correlations}, we first obtain the $y_i^h$ and $\hat{y}_i^h$ values. Since this time data is correlated, watermarks should be inserted in such a way that no malicious SP can understand the watermark inserted data points by checking the validity of the correlations. To guarantee this, if a data point $x_i$'s state is changed from $d_\alpha$ to $d_\beta$ (due to added watermark), the states of other data points that are correlated with $d_\beta$ state of $x_i$ should be also changed. Assume data has been shared for $h$ times before. Watermark insertion algorithm for the $(h+1)$-th sharing of the data with SP $\psi$ is described in Algorithms~\ref{alg:algo_main} and~\ref{alg:algo_recursive}.

\begin{algorithm}
		\KwIn{\\$\mathrm{Y}$=\{$y_0^h,\cdots,y_h^h$\}\\
			$D_{I_{\psi}}$ = \{$x_{1},\cdots,x_{\ell}$\}}
		\KwOut{$W_{I_{\psi}}$}
		\For{$t$ = 0 to $h$}{
			$\mathrm{T}_t$ = set of data points that are watermarked $t$ times\\
			sort $\mathrm{T}_t$ based on the presence probabilities\\
			\For{each $x_j \in \mathrm{T}_t$}{
				$d_j^*$ = value that maximizes presence probability of $x_j$\\
				insertWatermark($x_j$, $d_j^*$)\\
			}
		}
		\caption{ }
		\label{alg:algo_main}
\end{algorithm}
	
\begin{algorithm}[h]
		\KwIn{\\$x_j$ = data point to be watermarked\\
			$d_j^*$ = new value of $x_j$}
		\SetKwProg{Fn}{Function}{}{end}
		\Fn{$\mathrm{insertWatermark}$($x_j$, $d_j^*$)}{
			t = \# of times $x_j$ is watermarked during previous $h$ sharings\\
			\If{$\mathrm{Y[t]}$ $\neq$ 0 and $D_{I_\psi}$$\mathrm{[j]}$ $\neq$ ${\hat d}_j$}{
				$D_{I_\psi}$[j] = ${\hat d}_j$\\
				Y[t] - -\\
				$\mathrm{C}$ = set of data points correlated with ${\hat d}_j$ state of $x_j$\\
				\For{each $x_c \in \mathrm{C}$}{
					$d_c^*$ = desired value of $x_c$\\
					insertWatermark($x_c$, $d_c^*$)
				}
			}
		}
		\caption{ }
		\label{alg:algo_recursive}
\end{algorithm}
	
From the solution of the optimization problem, we know the number of data points which are watermarked $i$ times and will be watermarked in the current sharing ($y_i^h$). Since a data point could be watermarked between 0 and $h$ times, we have the solution set of the optimization problem as $Y$ = \{$y_0^h$, $y_1^h$, $\cdots$, $y_h^h$\}. Data points to be shared with SP $\psi$ are $D_{I_{\psi}}$ = \{$x_{1},\cdots,x_{\ell}$\} and the states of a data point are from the set $\{d_1, d_2, \cdots, d_m\}$. To add watermarks into data points that are watermarked for $t$ times ($t = 0, 1, \cdots, h$) in the previous $h$ sharings, we find the set of $t$ times watermarked data points ($\mathrm{T}_t$) and sort them in ascending order according to their presence probabilities (Algorithm 1, Line 2-3). Presence probability can be found as follows. Assume $d_j$ state of data point $x_j$ is correlated with the set of data points in $\mathrm{C} = \{x_{i_0} = d_{i_0}, \cdots, x_{i_n} = d_{i_n}$\}. Then, the presence probability for ($x_j=d_j$) is computed as $\prod_{t=0}^{n} Pr(x_j = d_j | x_{i_t} = d_{i_t})$.

Then, starting from the data point with minimum presence probability ($x_j$) in $\mathrm{T}_0$, we determine the state ($d_j^*$) that maximizes its presence probability and change the state of $x_j$ accordingly (Algorithm 1, Line 4-7). This way, we choose the most likely state value for $x_j$ according to the whole data. If the state of $x_j$ is already $d_j^*$, we skip this data point and continue with the next data point with minimum presence probability. Otherwise, we change the state of $x_j$ to $d_j^*$. Since we change a data point that is watermarked for $t$ times, we also decrement the value of Y[t] (=\:$y_t^h$) by 1.\footnote{Note that if Y[t] = 0, we skip the remaining $t$ times watermarked data points and repeat the same procedure for the data points in $\mathrm{T}_1$.} After the state of $x_j$ is changed to $d_j^*$, we find the data points that are correlated with $d_j^*$ state of $x_j$. That is, we construct a set $\mathrm{C}$ with data points that satisfy $Pr(x_i | x_j = d_j^*)>\tau$ and change the states of the data points in $\mathrm{C}$ (Algorithm 2, Line 6). For each data point in $\mathrm{C}$, we find its ``desired state'' (i.e., correlated state with $d_j^*$ state of $x_j$), and change it accordingly (Algorithm 2, Line 7-11). During this process, if we change a data point that is watermarked for $t^*$ times, we also decrement the value of Y[$t^*$] (=\:$y_{t^*}^h$) by 1. We continue this process until we add $w$ watermarks to the data. For some data, this algorithm may not find $w$ data points to add watermarks. For example, all data points may be in a state that maximizes its presence probability, and thus we may not find any data points to add watermark. In this case, instead of choosing the state for the highest presence probability, we choose the one for the second highest presence probability.

\begin{figure}[h!]
\centering
		\includegraphics[width=0.5\textwidth]{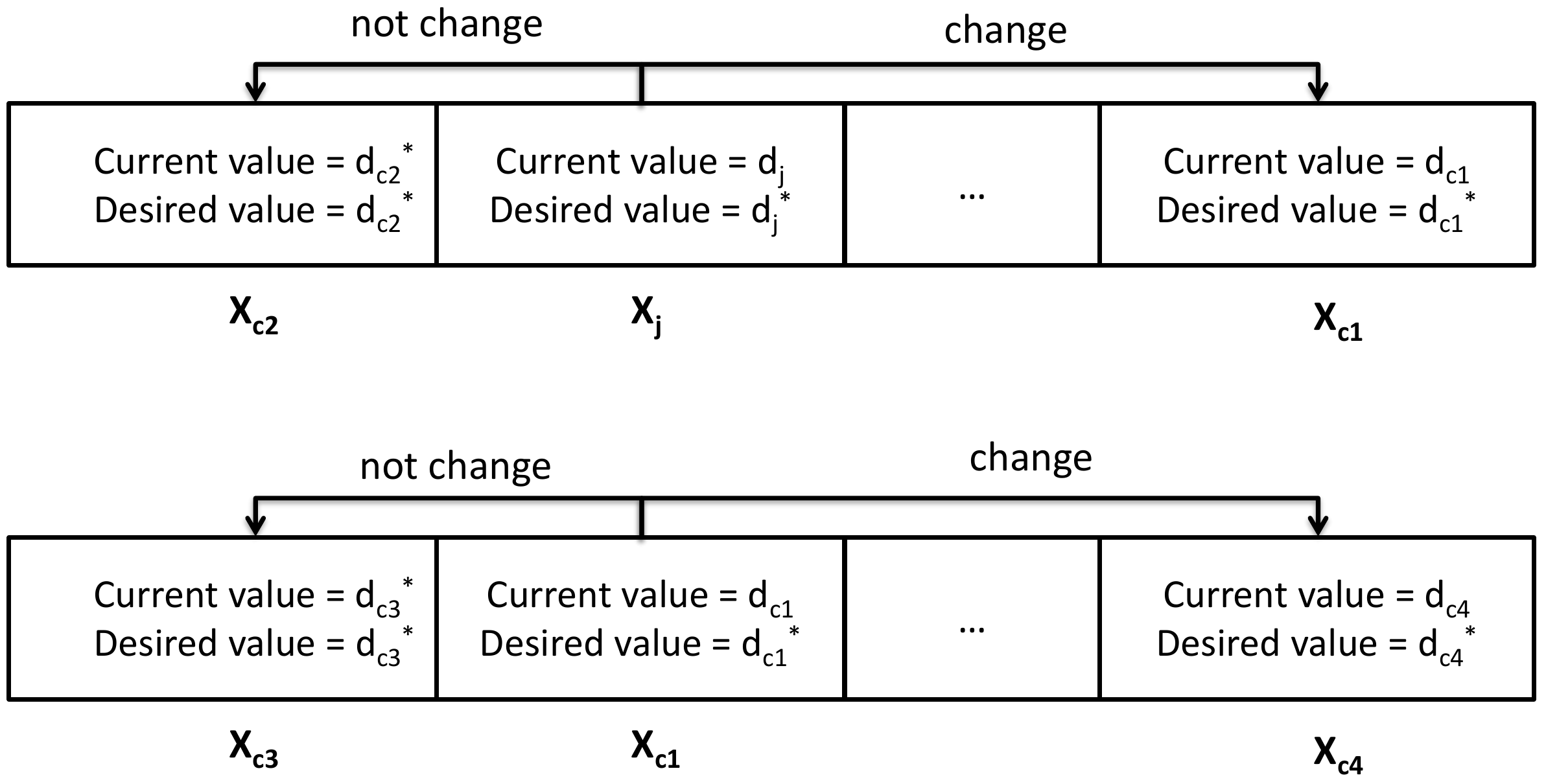}
		\caption{Toy example for inserting watermark into correlated data.}
		\label{fig:correlations}
\end{figure}

In Figure~\ref{fig:correlations}, we show this process with a small example. First, the state of data point $x_j$ is changed from $d_j$ to $d_j^*$. Then, data points ($x_{c1}$ and $x_{c2}$) which are correlated with $d_j^*$ state of $x_j$ are considered. Assume that $d_{c1}^*$ and $d_{c2}^*$ states of data points $x_{c1}$ and $x_{c2}$ are correlated with $d_j^*$ state of $x_j$. Data point $x_{c2}$ is already in state $d_{c2}^*$, thus we do not change its state. However, data point $x_{c1}$ is in state $d_{c1}$, and hence we change its state to $d_{c1}^*$. Since we change the state of $x_{c1}$ from $d_{c1}$ to $d_{c1}^*$, we now need to consider the data points that are correlated with $d_{c1}^*$ state of $x_{c1}$. Assume $d_{c3}^*$ and $d_{c4}^*$ states of data points $x_{c3}$ and $x_{c4}$ are correlated with $d_{c1}^*$ state of $x_{c1}$. Then, the state of $x_{c4}$ is changed since it does not have the desired value, but the state of data point $x_{c3}$ remains the same. This procedure continues with the data points which are correlated with $d_{c4}^*$ state of $x_{c4}$, until pre-defined watermark number is reached.
	
In this algorithm, we consider pairwise correlations between the data points. When correlations between the data points are more complex (e.g., higher order), we can still use a similar algorithm to handle them. We assume that malicious SPs also have the same resources we use in this algorithm to use the correlations (in order to detect the watermarked points) and evaluate the scheme accordingly in Section~\ref{sec:Evaluation}.

\color{black}
		
\section{Evaluation} \label{sec:Evaluation}

We implemented the proposed watermarking scheme on genomic data and evaluated its security (robustness) and utility guarantees. In this section, we provide the details of the data model we used in our evaluation and our results.

\subsection{Data Model}

For the evaluation, we used single-nucleotide polymorphism (SNP) data on the DNA. A SNP is a point variation on the DNA  that occurs when a single nucleotide adenine (A), thymine (T), cytosine (C), or guanine (G) in the genome differs between members of a species~\cite{snp}. For example, two sequenced DNA fragments from different individuals, AAGCCTA and AAGCTTA, contain a difference in a single nucleotide (at the $5^{th}$ position). In this case, we say that there are two alleles: C and T. Almost all common SNPs have only two alleles, and everyone inherits one allele of every SNP position from each of his parents. If an individual receives the same allele from both parents, he is said to be homozygous for that SNP position. If however, he inherits a different allele from each parent (one minor and one major), he is called heterozygous. Depending on the alleles the individual inherits from his parents, the state (or value) of a SNP position can be simply represented as the number of minor alleles it possesses, i.e., 0, 1, or 2. We obtained SNP data of 99 individuals from 1000 Genomes Project~\cite{1000gp}. In the obtained dataset, each individual has 7690 SNP values meaning that we have a 99 by 7690 matrix and elements of matrix are either 0, 1, or 2.

\subsection{Results}

We evaluated the proposed watermarking scheme in various aspects. In particular, we evaluated its security (robustness) against collision and correlation attacks (as discussed in Section~\ref{sec:The Threat Model}) and the loss in data utility due to watermark addition. In all collusion attack scenarios, we assume that Alice shares the same data portion with the SPs. This assumption provides the maximum amount of information to the malicious SPs. If different set of data points are shared with the SPs, malicious SPs can use the intersection of these data points for the collusion attack. We also evaluated the proposed scheme in terms of the (watermark) detection performance of the data owner under various attacks. We ran all experiments for $1000$ times and report the average values.

\subsubsection{Robustness against watermark inference}

Here, we evaluate the robustness of the proposed scheme against watermark inference.

\begin{figure}[h!]
\centering
	\includegraphics[width=0.5\textwidth]{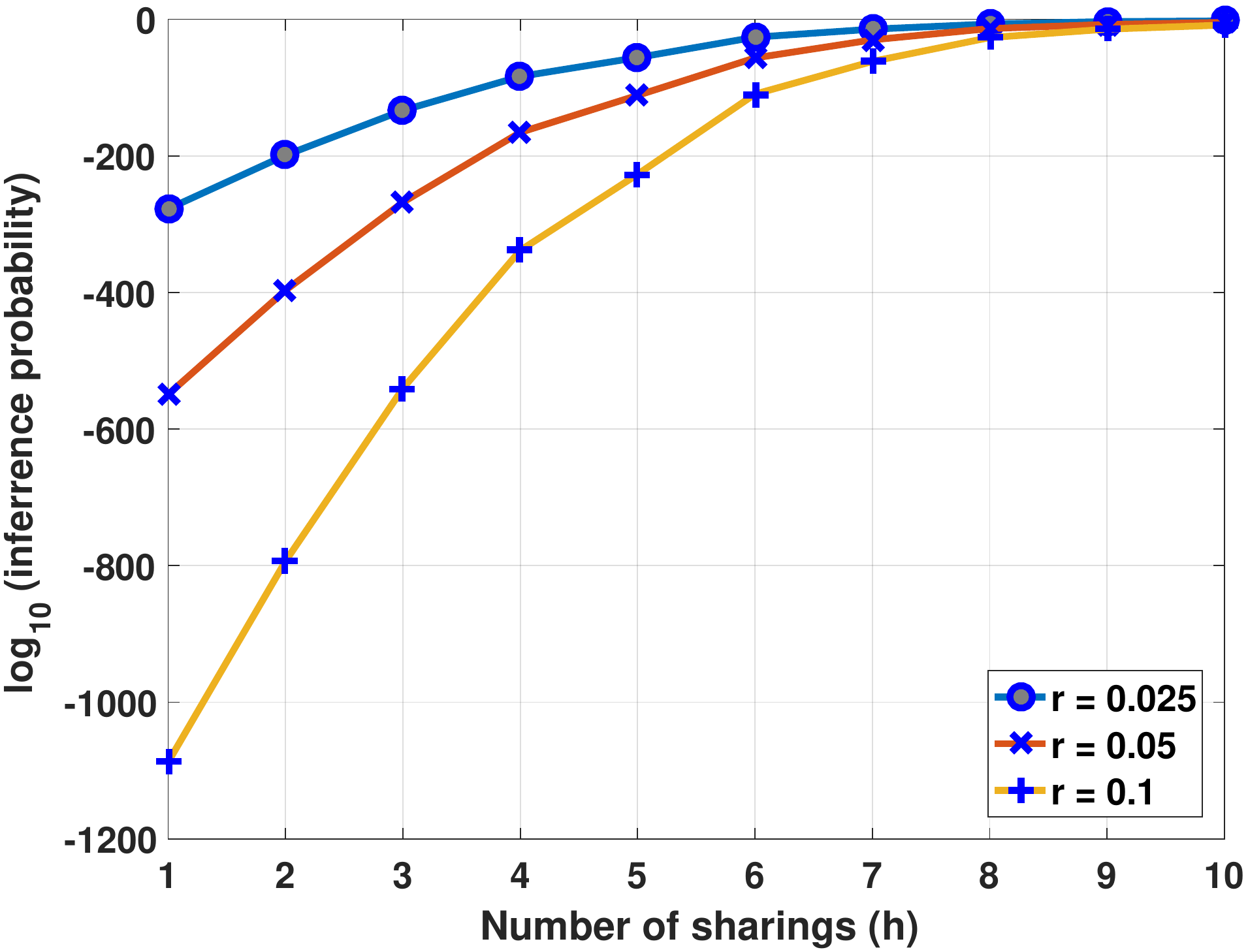}
	\caption{Probability of identifying the whole watermarked points in the collusion attack when $h$ malicious SPs collude. $r$ represents the fraction of watermarked data.}
	\label{fig:plot_prob_wat_len}
\end{figure}

\noindent\textbf{Collusion attack:} First, we evaluated the probability of identifying the whole watermarked points in the collusion attack (when correlations in data are not considered). We considered the worst case scenario and assumed that all the SPs that has Alice's data are malicious, and hence they exactly know how many times Alice has shared her data to compute the exact probabilities for the attack (as discussed in Section~\ref{sec:For Data With No Correlations}). In Figure~\ref{fig:plot_prob_wat_len}, we show the logarithm of this inference probability when data is shared with $h$ SPs and they are all malicious (where $h=(1,2,\cdots,10)$) and when different fractions of data is watermarked. Assuming Alice's shared data is of length $\ell$ and the length of the added watermark is $w$, we denote the fraction of watermarked data (or watermark ratio) as $r=w/\ell$. Detailed results of this experiment are also shown in Table~\ref{table:table_prob_wat_len}. Overall, we observed that the probability to completely identify the watermark via the collusion attack is significantly low when the proposed technique is used for watermarking the data. Following our definition of robustness against watermark inference (in Section~\ref{sec:robustness}), under this attack model, the proposed scheme is $p$-robust against $f$-watermark inference for $f=1$ and $p\leq10^{-2}$ when $h$ is as high as $10$ and data utility is as high as $97\%$ (i.e., $r$ is as small as $0.025$). As expected, we observed that the inference probability of the malicious SPs increases with decreasing $r$ and increasing $h$ values. \color{black} That is, as data is shared with more malicious SP, the probability to identify the watermarked data increases due to the collusion attack. \color{black} Also note that even for significantly low values of $r$ (that corresponds to high data utility), the proposed scheme provides high resiliency against collusion attacks.

\begin{table}[h!]
	\centering
	\begin{tabular}{|c|l|l|l|l|l|}
		\hline
		\textbf{} & \multicolumn{4}{c|}{\textbf{Watermark ratio ($r$)}} \\ \hline
		\multirow{7}{*}{\textbf{\begin{tabular}[c]{@{}c@{}}Number\\ of\\ sharings\\ ($h$)\\ \\ \end{tabular}}} &  & \textbf{0.025} & \textbf{0.05} & \textbf{0.1}  \\ \cline{2-5}
		& \textbf{2} & -199 & -397 & -793  \\ \cline{2-5}
		& \textbf{4} & -84 & -166 & -338  \\ \cline{2-5}
		& \textbf{6} & -26 & -57 & -110  \\ \cline{2-5}
		& \textbf{8} & -7 & -14 & -27  \\ \cline{2-5}
		& \textbf{10} & -2 & -5 & -8  \\ \hline
	\end{tabular}
	\caption{Common logarithm ($\log_{10}$) of the inference probability to identify the whole watermark for varying $h$ (number of colluding SPs) and $r$ (fraction of watermarked data) values.}
	\label{table:table_prob_wat_len}
\end{table}

\begin{figure}[h!]
\centering
	\includegraphics[width=0.5\textwidth]{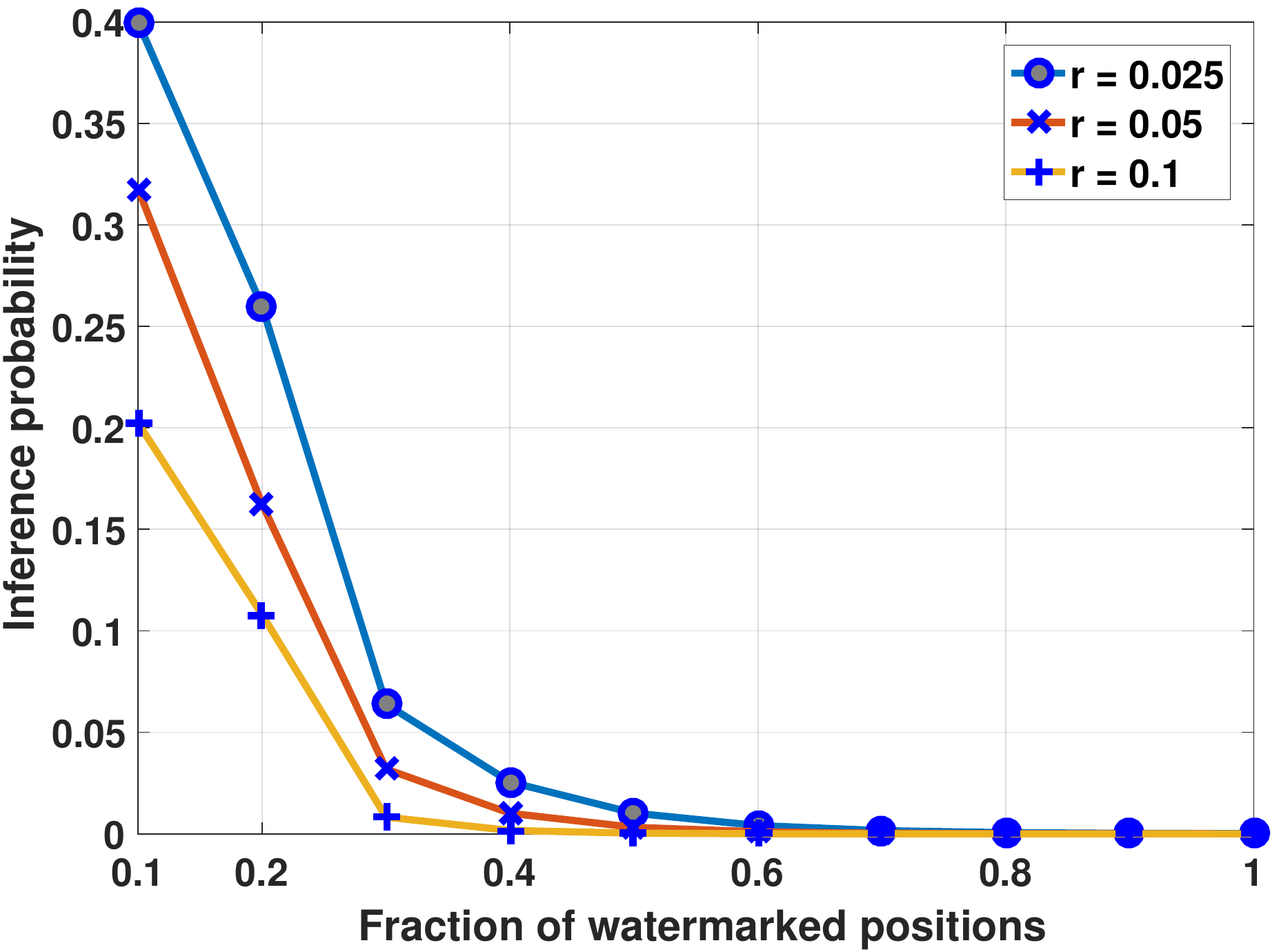}
	\caption{Inference probability to identify different fractions of the watermarked positions in the collusion attack when the number of colluding malicious SPs $h=6$. $r$ represents the fraction of watermarked data.}
	\label{fig:plot_ratio_prob}
\end{figure}

We also ran the same experiment to observe the probability of malicious SPs to identify different fractions of the watermarked positions. In Figure~\ref{fig:plot_ratio_prob}, we show this inference probability. For this experiment, we assume that the malicious SPs initially try to identify the watermark positions that has higher probability to be watermarked. Since we assume that the watermarking algorithm is publicly known by the malicious SPs, once they observe vertically aligned data points (as in Figure~\ref{fig:collusion_attack_1}), they can compute the probability of being watermarked for each data position (as discussed in Section~\ref{sec:For Data With No Correlations}) and initially try to identify high probability watermark positions. We also set the number of colluding malicious SPs $h=6$ and watermarked different fractions of the whole data (i.e., varied the $r$ value). We observed that colluding SPs can identify small portion of watermark locations with small probabilities and this probability rapidly decreases with increasing fraction of watermarked data ($r$). Also, the probability to identify more than $30\%$ of the watermarked locations is significantly low even when the malicious SPs collude. Notably, we show that when $r=0.025$ (which means 200 watermarked data points on a data of size 7690, and hence preserves more than $97\%$ of data utility), even when 6 malicious SPs collude, the probability to recover more than $30\%$ of the watermark locations is very small. In other words, under this attack model, when $r=0.025$, the proposed scheme is $p$-robust against $f$-watermark inference for $f=0.3$ and $p\leq10^{-1}$.

Next, we considered the case in which the malicious SPs do not exactly know how many times Alice has shared her data (i.e., malicious SPs with partial knowledge as discussed in Section~\ref{sec:For Data With No Correlations}). Thus, we evaluated the relation between the inference probability (to identify the whole watermarked positions) and the assumption of the malicious SPs on the number of times data has been shared. As discussed in Section~\ref{sec:For Data With No Correlations}, in practice, $h$ colluding SPs can only know that data has been previously shared for at least $h$ times and they make assumptions about the exact number of sharings. Hence, to compute the probabilities they use for the attack, they may assume that data has been shared for between $h$ and $(h+t)$ times, where $(h+t)$ is an upper limit.

\begin{figure}[h!]
\centering
	\subfloat[Different number of colluding SPs ($h$) when fraction of watermarked data $r=0.05$.]{\includegraphics[clip,width=0.75\columnwidth]{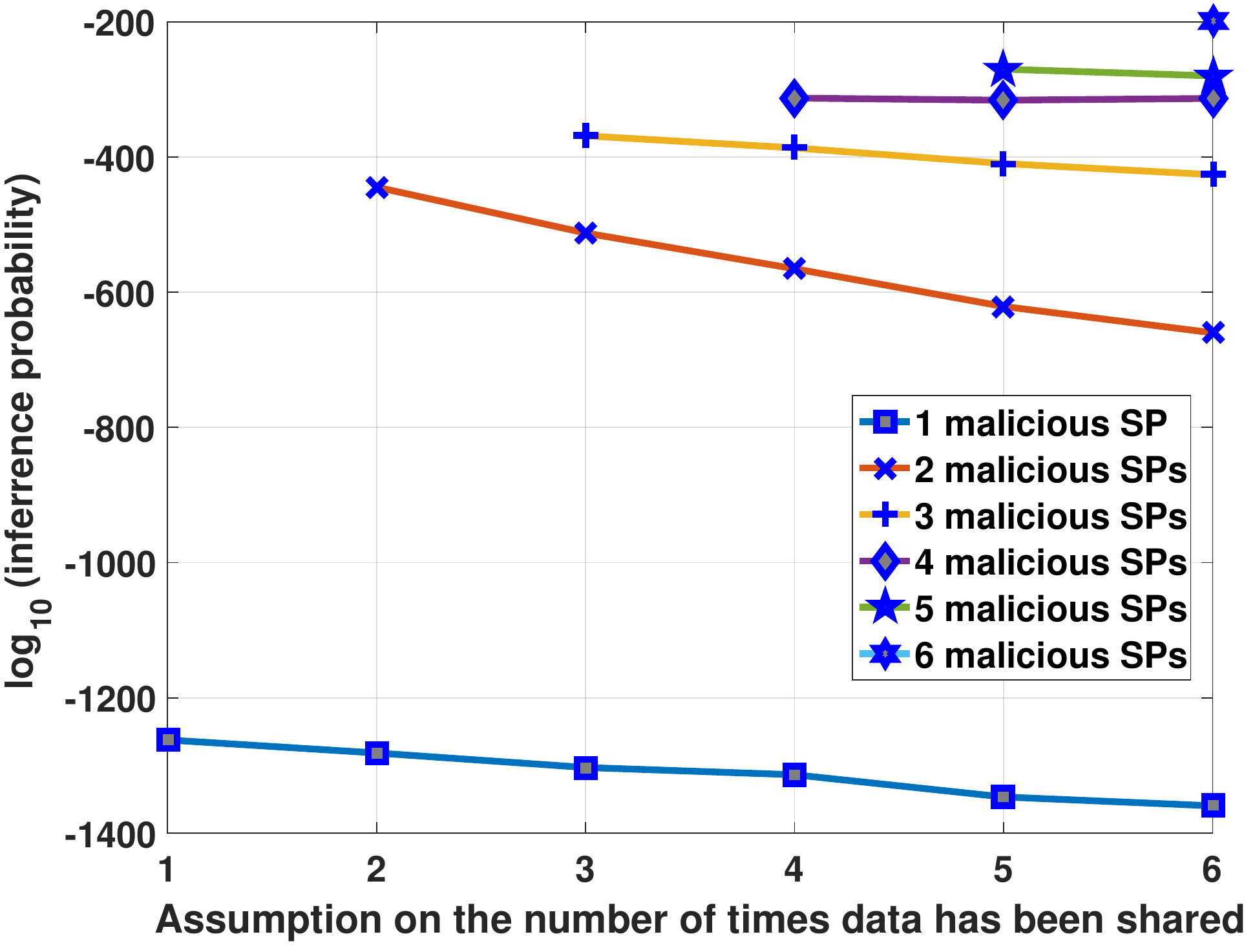}}
	
	\subfloat[Different fractions of watermarked data for a single malicious SP.]{\includegraphics[clip,width=0.75\columnwidth]{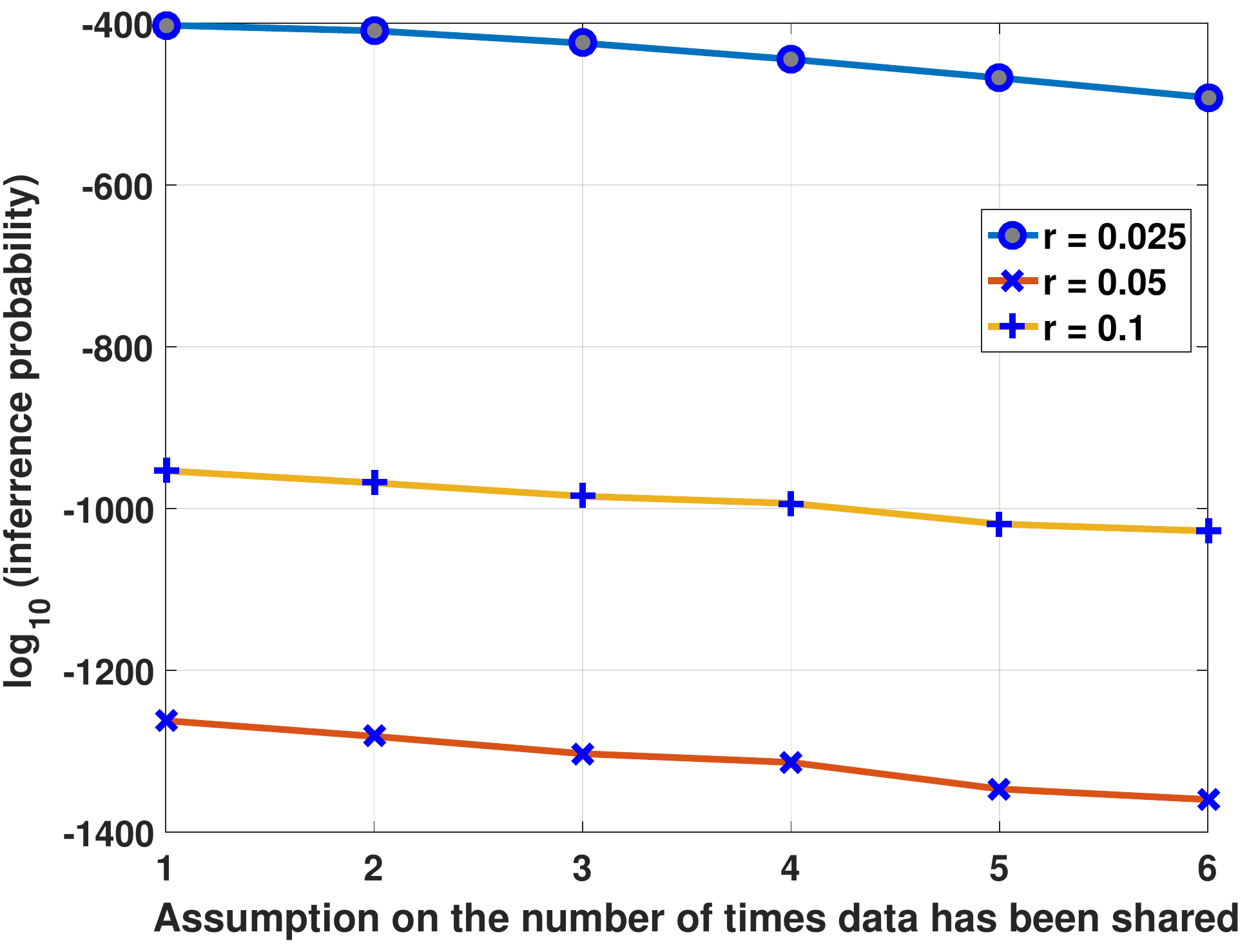}}
	\caption{Probability of identifying the whole watermarked points in the collusion attack when malicious SPs has partial knowledge about the number of times data has been shared. Data has been actually shared for 6 times.}
	\label{fig:plot_prob_assumption}
\end{figure}

In Figure~\ref{fig:plot_prob_assumption} we show the logarithm of the inference probability when data has been actually shared for 6 times by Alice. In Figure~\ref{fig:plot_prob_assumption}(a), we assume different number of colluding SPs ($h$) that run the collusion attack as discussed in Section~\ref{sec:For Data With No Correlations} (as malicious SPs with partial knowledge). For instance, when $h=3$, the malicious SPs run the attack four times assuming that data has been shared for 3, 4, 5, and 6 times, respectively. For this experiment, we set $r=0.05$. In Figure~\ref{fig:plot_prob_assumption}(b), we show the same probability for different $r$ values for a single malicious SP. We observed that as the colluding SPs infer more missing data points (even when they correctly guess the exact number of sharings), their inference probability decreases. Therefore, it is better for $h$ colluding SPs to assume that data has been shared for exactly $h$ times, and run their attack accordingly. However, even in this case, we show that the inference probability of the malicious SPs is significantly low.

\noindent\textbf{Correlation attack:} To evaluate the security of the proposed scheme against the correlations in the data, we compared two techniques presented in Sections~\ref{sec:For Data With No Correlations} and~\ref{sec:For Data With Correlations}. In this analysis, we focused on a data length ($\ell$) of 100 in our dataset. We find each pairwise correlation $Pr(x_i = \alpha | x_j = \beta)$ between these 100 data points, where $\alpha,\beta \in \{0, 1, 2\}$. To consider only strong correlations (and to avoid the noise that arise due to weak correlations), we only consider the ones above a threshold $\tau$ (we selected $\tau=0.9$). Note that the correlations in the data are not symmetric. That is, $Pr(x_i = d_i | x_j = d_j)$ being high does not mean that $Pr(x_j = d_j | x_i = d_i)$ is also high.

\begin{figure}[h!]
\centering
	\subfloat[Correlations in the data are not considered when selecting the data points to be watermarked (i.e., technique proposed in Section~\ref{sec:For Data With No Correlations} is used for watermarking).]{\includegraphics[clip,width=0.75\columnwidth]{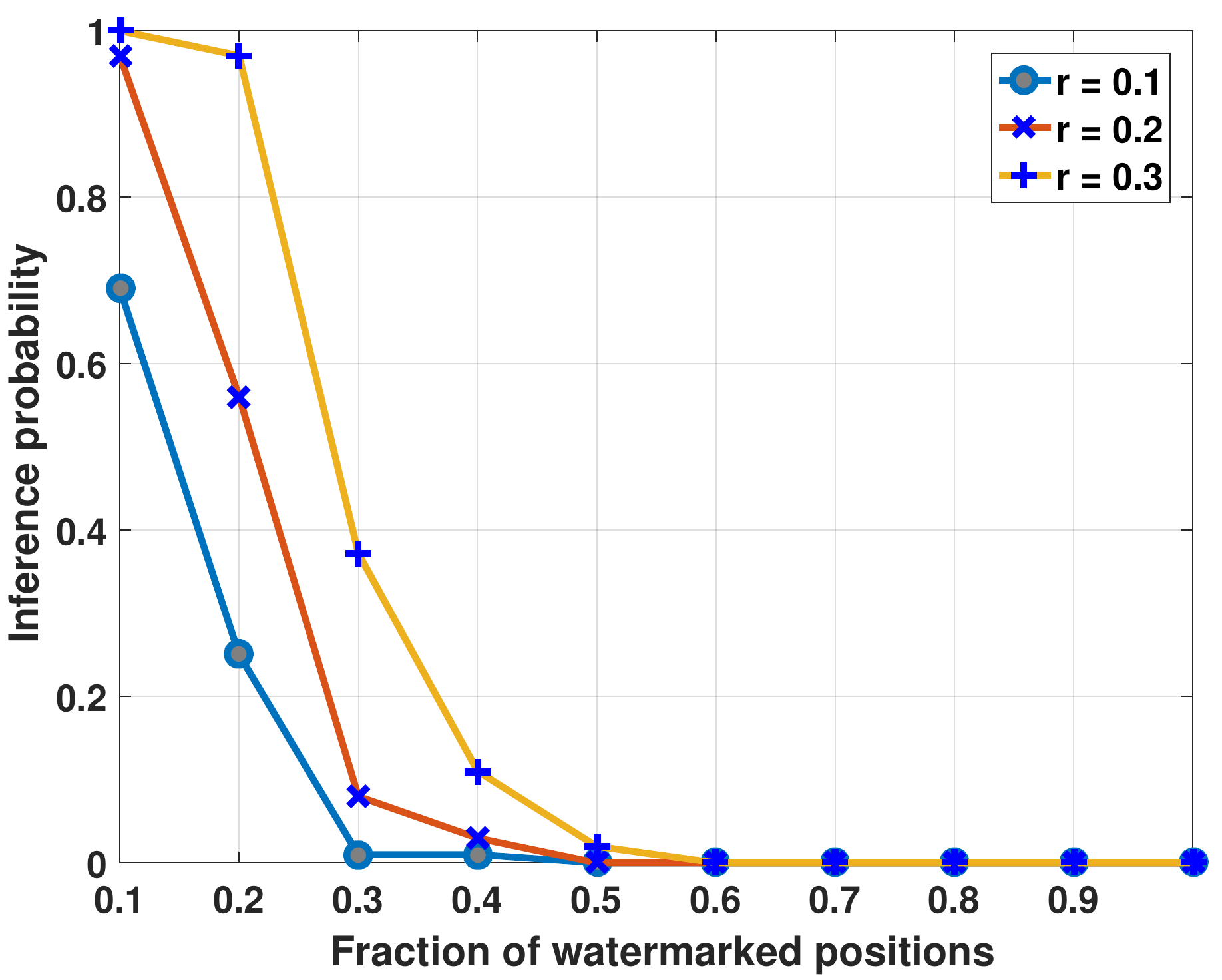}}
	
	\subfloat[Correlations in the data are considered using the proposed algorithm when selecting the data points to be watermarked (i.e., technique proposed in Section~\ref{sec:For Data With Correlations} is used for watermarking).]{\includegraphics[clip,width=0.75\columnwidth]{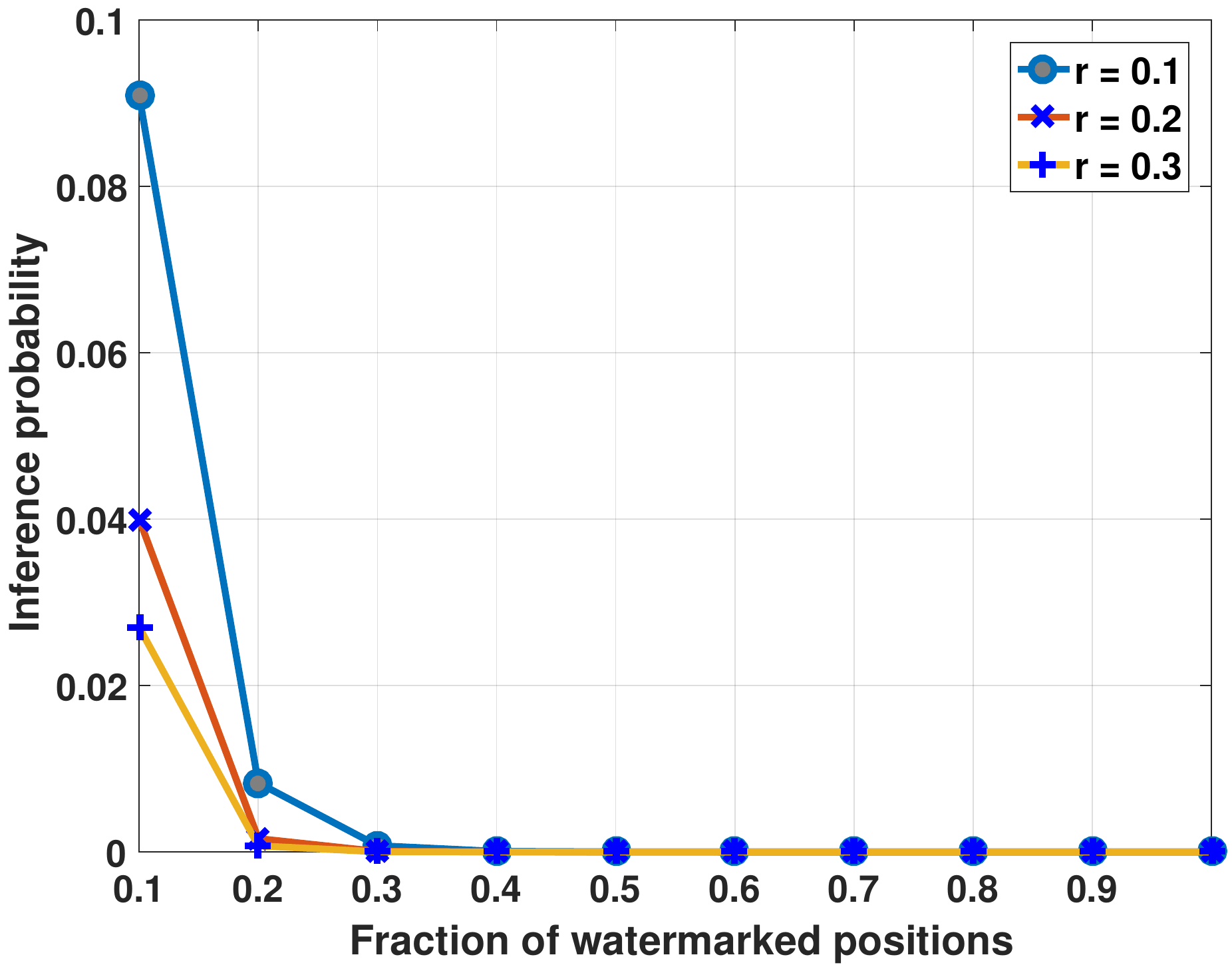}}
	\caption{Inference probability to identify different fractions of the watermarked positions in the single SP correlation attack. $r$ represents the fraction of watermarked data.}
	\label{fig:correlation}
\end{figure}

First, we compared two schemes for a single SP attack in terms of the probability of the malicious SP to identify different fractions of the watermarked positions. Note that in this attack, the malicious SP also utilizes its knowledge of correlations in the data.\footnote{We assume that knowledge of the malicious SP about the correlations is the same as the knowledge we utilized while adding the watermark in Section~\ref{sec:For Data With Correlations}.} In Figure~\ref{fig:correlation}(a) and (b) we show this comparison for different $r$ values. We observed in Figure~\ref{fig:correlation}(a) that as $r$ increases, the inference probability of the malicious SP increases for the technique presented in Section~\ref{sec:For Data With No Correlations}. This is expected since (i) if correlations are not considered while selecting the watermarked positions, the probability of the attacker to identify the watermarked positions also increases, and (ii) as more data points are watermarked in this way, the attacker can identify more watermarked position. However, when we consider the correlations in the data when selecting the watermark locations, the inference probability of the malicious SP significantly decreases as shown in Figure~\ref{fig:correlation}(b). Also, in this scenario, inference probability decreases with increasing $r$ value as expected. For instance, when $r=0.3$, the watermarking scheme is $p$-robust against $f$-watermark inference for $f=0.2$ and $p\simeq1$ when the correlations in the data are not considered. When we consider the correlations in the data using the proposed watermarking algorithm, it becomes $p$-robust against $f$-watermark inference for $f=0.2$ and $p\simeq0$.

\noindent\textbf{Collusion and correlation attack:} We also compared two techniques presented in Sections~\ref{sec:For Data With No Correlations} and~\ref{sec:For Data With Correlations} to show the resiliency of the proposed watermarking scheme against both collusion and correlation attacks at the same time. In this attack, each malicious SP first runs the correlation attack independently. As a result of this part, each malicious SP detects a number of watermarked points. For the advantage of the malicious SPs (and to consider the worth case scenario), we consider the outcome of the malicious SP with the highest number of correct detections. Let the number of watermarks detected by this malicious SP be $m$ as a result of the first part.\footnote{As discussed, malicious SPs may detect less than $w$ watermarked points as a results of the correlation attack.} Then, to detect the remaining $w-m$ watermarked points, malicious SPs run the collusion attack.

\begin{figure}[h!]
\centering
	\subfloat[Correlations in the data are not considered when selecting the data points to be watermarked (i.e., technique proposed in Section~\ref{sec:For Data With No Correlations} is used for watermarking).]{\includegraphics[clip,width=0.75\columnwidth]{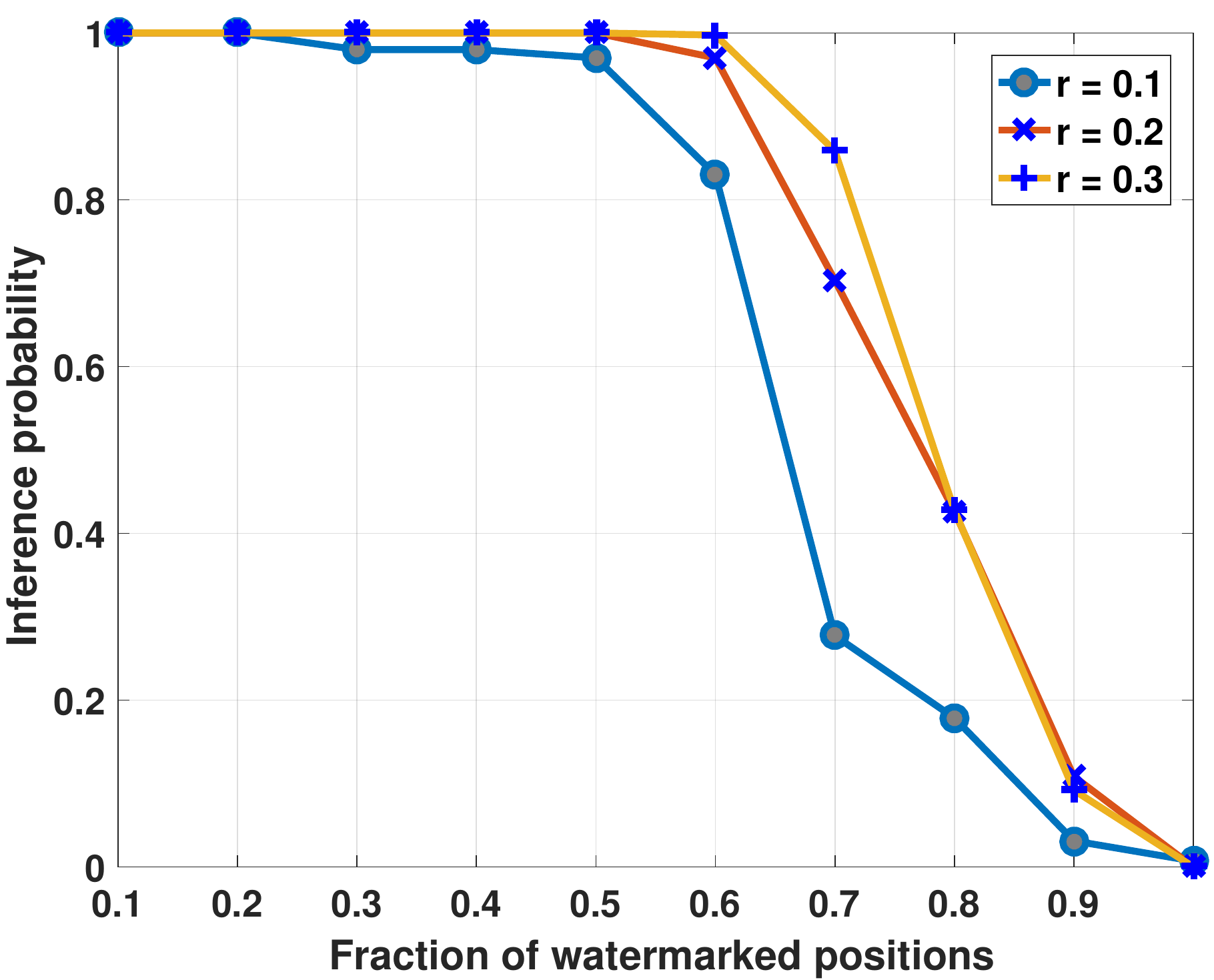}}
	
	\subfloat[Correlations in the data are considered using the proposed algorithm when selecting the data points to be watermarked (i.e., technique proposed in Section~\ref{sec:For Data With Correlations} is used for watermarking).]{\includegraphics[clip,width=0.75\columnwidth]{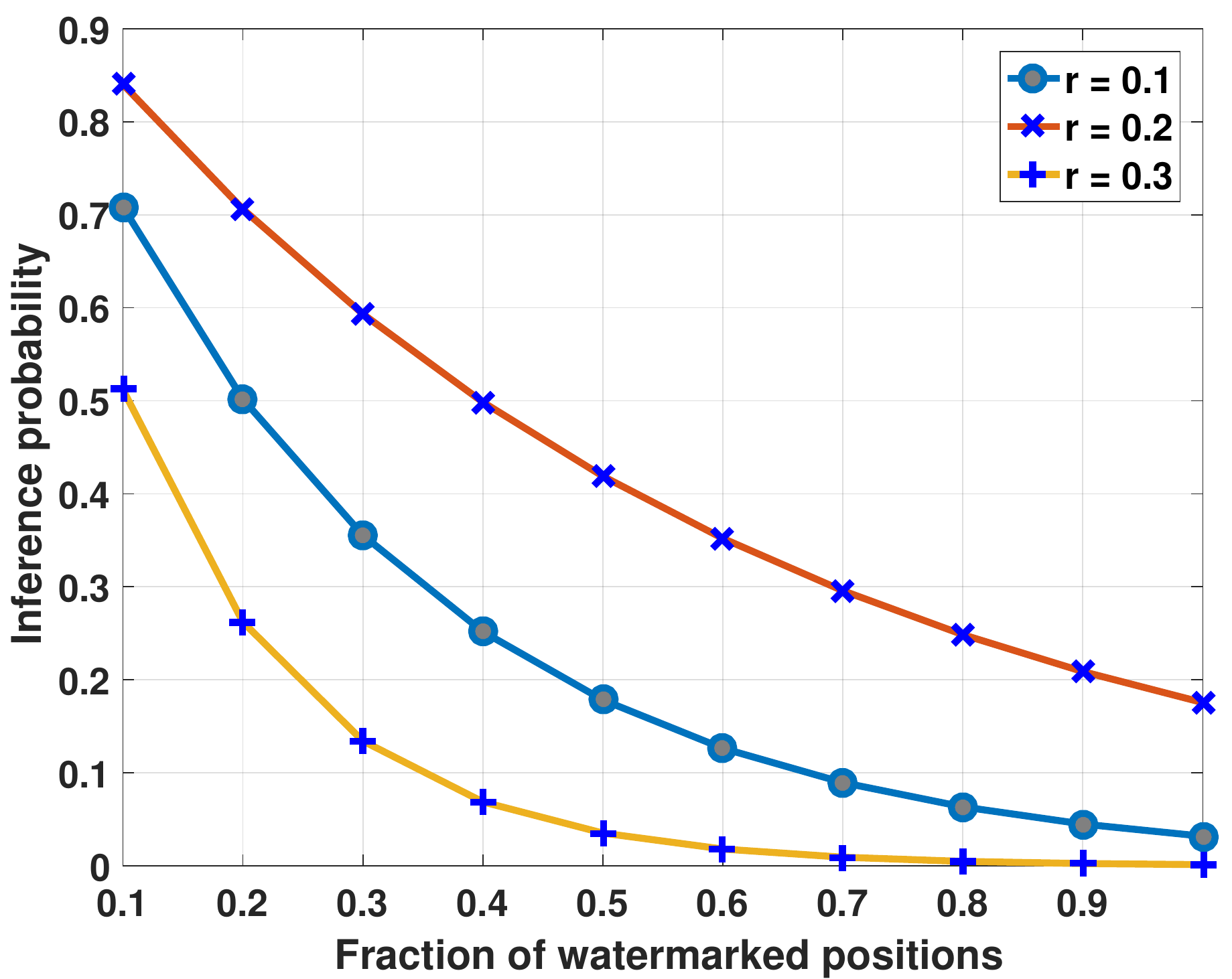}}
	\caption{Inference probability to identify different fractions of the watermarked positions in collusion attack (when $h=6$) in which the malicious SPs also use the correlations in the data. $r$ represents the fraction of watermarked data.}
	\label{fig:collusion_correlation}
\end{figure}

In Figure~\ref{fig:collusion_correlation}(a) and (b) we show this comparison for different $r$ values when the number of colluding malicious SPs $h=6$ (and data has been shared for $6$ times). We observed that when the correlations are not considered in the watermarking algorithm, malicious SPs can identify more than half of the watermarked data locations with high probability as shown in Figure~\ref{fig:collusion_correlation}(a). However, when we consider the correlations to select the data points to be watermarked, the inference probability of the malicious SPs significantly decreases (as in Figure~\ref{fig:collusion_correlation}(b)). For instance, when $r=0.3$, the watermarking scheme is $p$-robust against $f$-watermark inference for $f=0.5$ and $p\simeq1$ when the correlations in the data are not considered. When we consider the correlations in the data using the proposed watermarking algorithm, it becomes $p$-robust against $f$-watermark inference for $f=0.5$ and $p\leq0.1$. This shows that the proposed watermarking scheme provides security guarantees against both collusion and correlation attacks with high probabilities even when all the SPs that receive the data are malicious and colluding (as in this experiment). Note that in Figure~\ref{fig:collusion_correlation}(b), the reason inference probabilities for $r=0.2$ is larger than the ones for $r=0.1$ is due to the result of the optimization problem.

\color{black}

\subsubsection{Robustness against watermark modification}\label{sec:detection_performance}

Here, we evaluate the robustness of the proposed scheme against watermark modification.

\noindent\textbf{Partial sharing:} We evaluated the detection performance (and robustness against watermark modification) of the proposed watermarking scheme when a malicious SP partially shares Alice's data. In this scenario, we assume that Alice has shared her data (same data portion at each sharing) with $h$ SPs ($SP_1,\cdots, SP_h$). The malicious SP, rather than sharing the whole data with a third party without Alice's authorization, shares different fractions of the data to avoid being detected by Alice. As we have shown in previous experiments, the probability for a malicious SP to detect the watermarked data points is significantly low for our proposed scheme (even in the existence of collusion attack). Thus, we assume that the malicious SP randomly selects different fractions of data points to share with the third party. Here, we assume the malicious SP does not further modify Alice's data before it shares it with a third party as such modification would degrade the credibility of the data (as we discuss in Section~\ref{sec:discussion}). We also consider and extensively study the impact of such modification to the detection performance later in this section.

\begin{figure}[h!]
\centering
	\subfloat[Different fractions of watermarked data ($r$) when data has been shared with $h=4$ SPs.]{\includegraphics[clip,width=0.75\columnwidth]{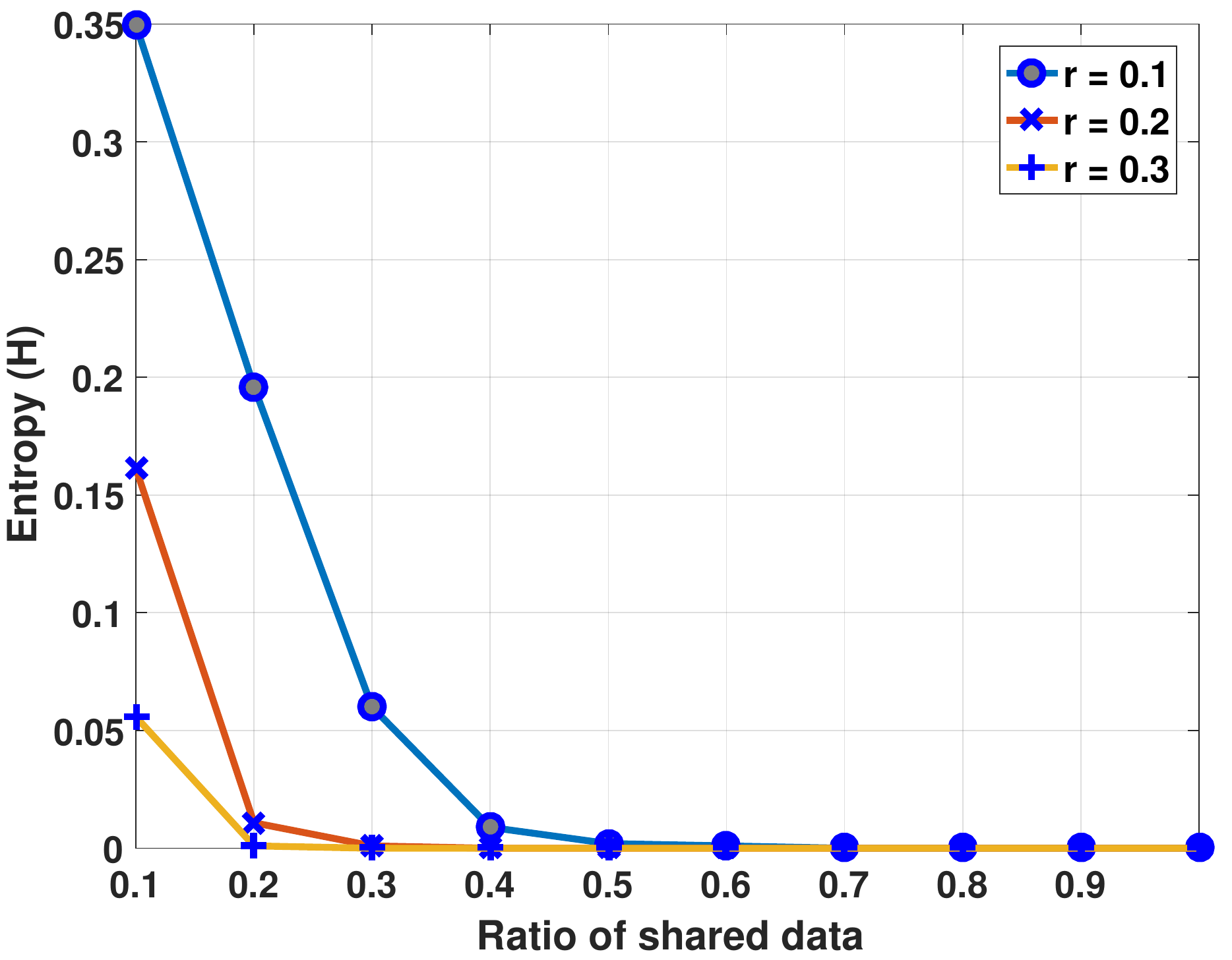}}
	
	\subfloat[Alice shares her data with $h$ SPs when fraction of watermarked data $r=0.2$.]{\includegraphics[clip,width=0.75\columnwidth]{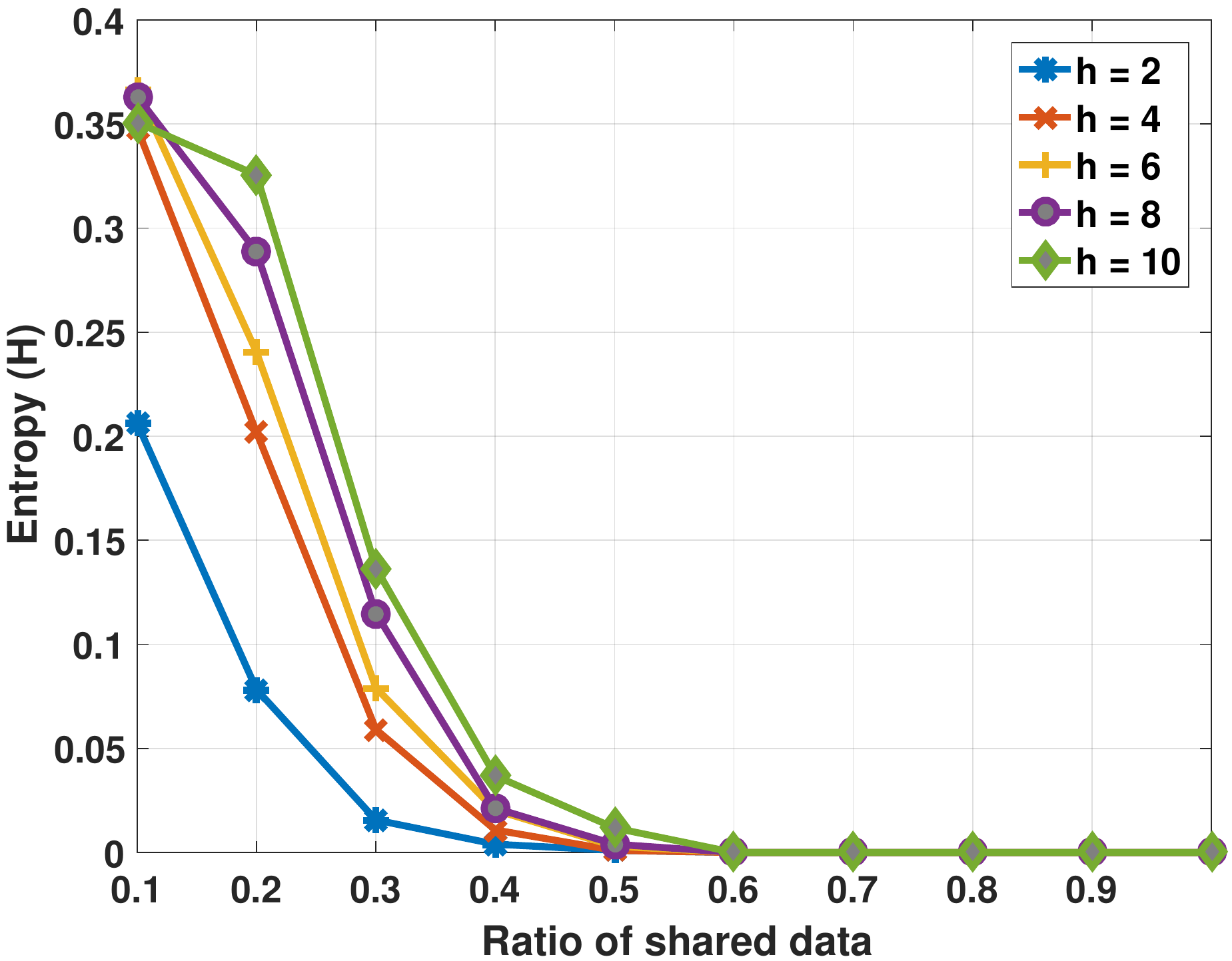}}
	\caption{Uncertainty of the data owner (Alice) to identify the source of the data leakage when the malicious SP partially shares Alice's data.}
	\label{fig:entropy}
\end{figure}

We first quantify the detection performance of our proposed scheme under this scenario by using an entropy metric. That is, we compute the uncertainty of Alice about the source of the leaked data. Alice, to detect the source of the leak, compares the watermark pattern corresponding to each SP she has shared her data with the (partial) pattern on the leaked data. Thus, any SP $i$ whose watermark pattern ($Z_{I_i}$) includes the partial watermark is marked as a potential malicious SP by Alice. Let $SP_\phi$ be the source of the leak as inferred by Alice as a result of this comparison. The probability that the source being $SP_i$ is then $p(SP_\phi=SP_i)$. Thus, we compute the uncertainty of Alice about the source of the leak as $H=-\sum_{i=1}^hp(SP_\phi=SP_i)\log{p(SP_\phi=SP_i)}$. We show the results of this evaluation in Figures~\ref{fig:entropy}(a) and (b) for varying $r$ and $h$ values, respectively. We conclude that the data owner can associate the source of the leakage to the corresponding SP with high probability in most of the cases, except when the malicious SP shares very small portion of user's data with a third party. However, this particular case would also reduce the benefit of the malicious SP (due to the unauthorized sharing) significantly. Furthermore, such partial sharing may degrade the credibility of data as discussed in Section~\ref{sec:discussion}.

\begin{figure}[h!]
\centering
	\subfloat[Different fractions of watermarked data ($r$) when data has been shared with $h=4$ SPs.]{\includegraphics[clip,width=0.75\columnwidth]{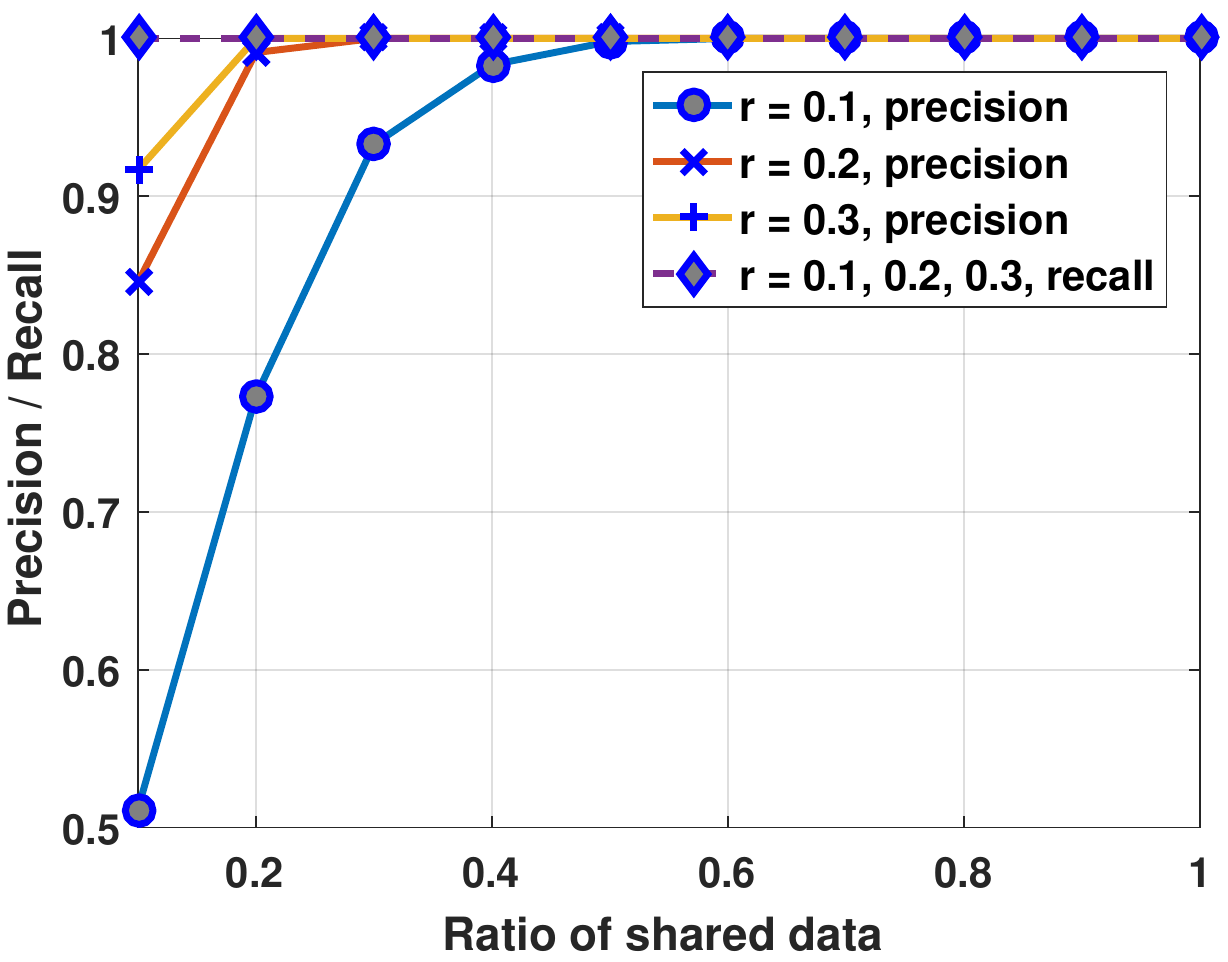}}
	
	\subfloat[Alice shares her data with $h$ SPs when fraction of watermarked data $r=0.2$.]{\includegraphics[clip,width=0.75\columnwidth]{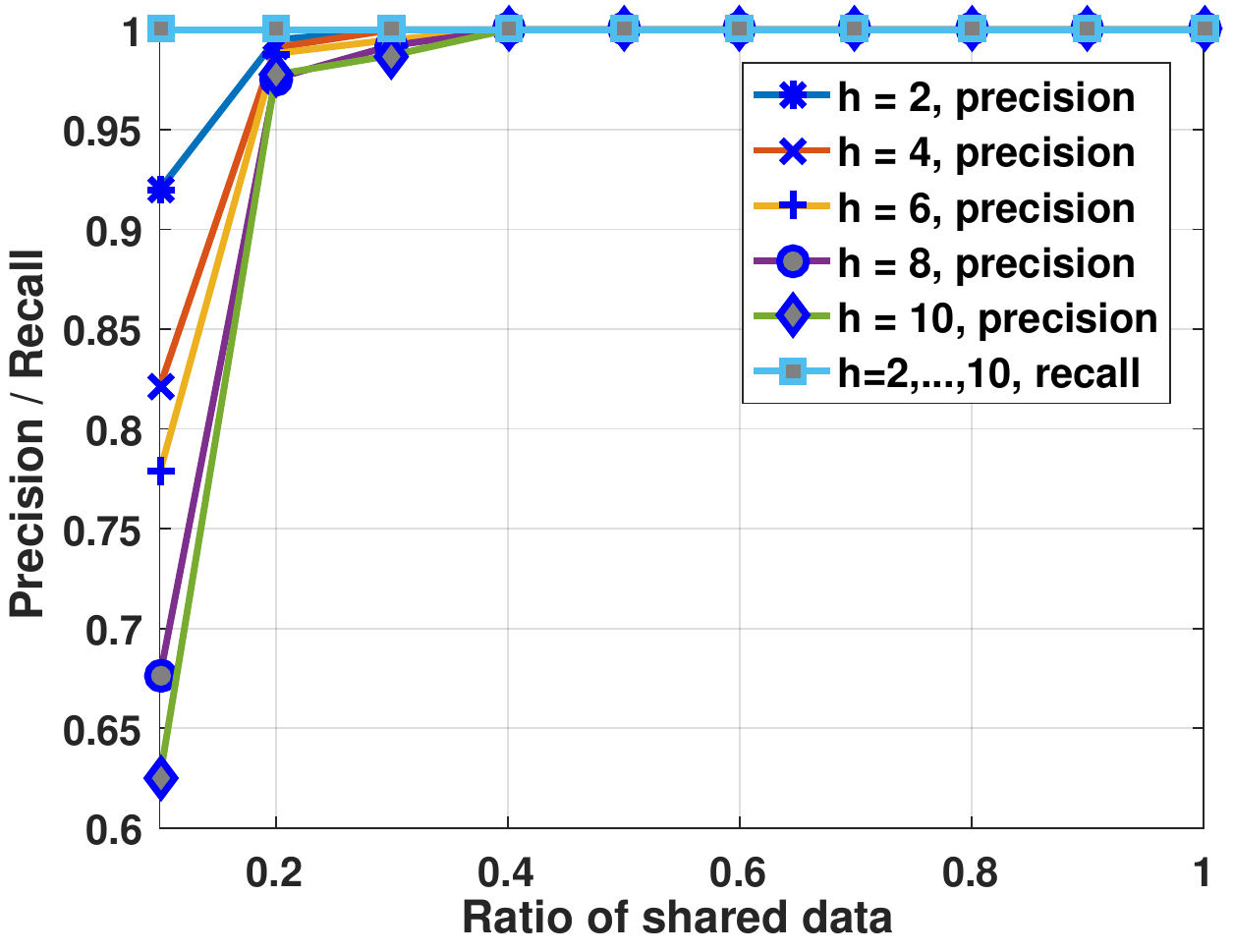}}
	\caption{Precision and recall values for the data owner to detect the malicious SP when the malicious SP partially shares Alice's data.}
	\label{fig:precisionrecall}
\end{figure}

We also quantify the robustness against watermark modification under this attack using precision and recall metrics. Alice constructs a set $\mathbf{S}$ that includes the malicious SPs detected by her. We define true positive as a malicious SP that is in set $\mathbf{S}$, false positive as a non-malicious SP that is in $\mathbf{S}$, true negative as a non-malicious SP that is not in $\mathbf{S}$, and false negative as a malicious SP that is not in $\mathbf{S}$. In Figures~\ref{fig:precisionrecall}(a) and~(b), we show the precision and recall values for varying ratio of shared data by the malicious SP and for different $r$ and $h$ values. Following our definition of robustness against watermark modification (in Section~\ref{sec:robustness}), under this attack model, the proposed scheme is $\rho/\epsilon$-robust against watermark modification with $\epsilon=1$ for all considered values of $h$ and $r$. Furthermore, when $r\geq0.2$, $h\leq10$, and the ratio of shared data by the malicious SP is more than $0.2$, the proposed scheme is $\rho/\epsilon$-robust against watermark modification with $\rho\simeq0.97$ and $\epsilon=1$.


\noindent\textbf{Watermark modification:} Finally, we studied a stronger attack in which malicious SP (or SPs) modify the data in order to damage the watermark (and hence, it becomes harder for the data owner to detect the source of the data leak). Note that in practice, such modification of data not only reduces data utility (as we show in our experiments), but it also degrades data credibility while the malicious SPs share the data with a third party (as discussed in Section~\ref{sec:discussion}). Here, malicious SPs (or SP) try to remove or damage the watermark by (i) changing the states of data points that are different when they aggregate their data (i.e., when they detect a data point with multiple states in the aggregate data, they change its state to the majority of the observed states), and (ii) adding noise to other data points (i.e., changing states of other random data points). Eventually, data leaked by the malicious SPs has a watermark pattern represented as $Z_\alpha$. Using $Z_\alpha$ and unique watermark patterns of the SPs (that previously received the data), Alice constructs the set $\mathbf{S}$ that includes the malicious SPs detected by her. As before, we evaluate the success of the detection via precision and recall metrics. For all following experiments we set the watermark ratio ($r$) to $0.05$.

\begin{figure}[h!]
\centering
	\subfloat[Data owner knows the number of malicious SPs ($\phi=\hat{\phi}=1$).]{\includegraphics[clip,width=0.75\columnwidth]{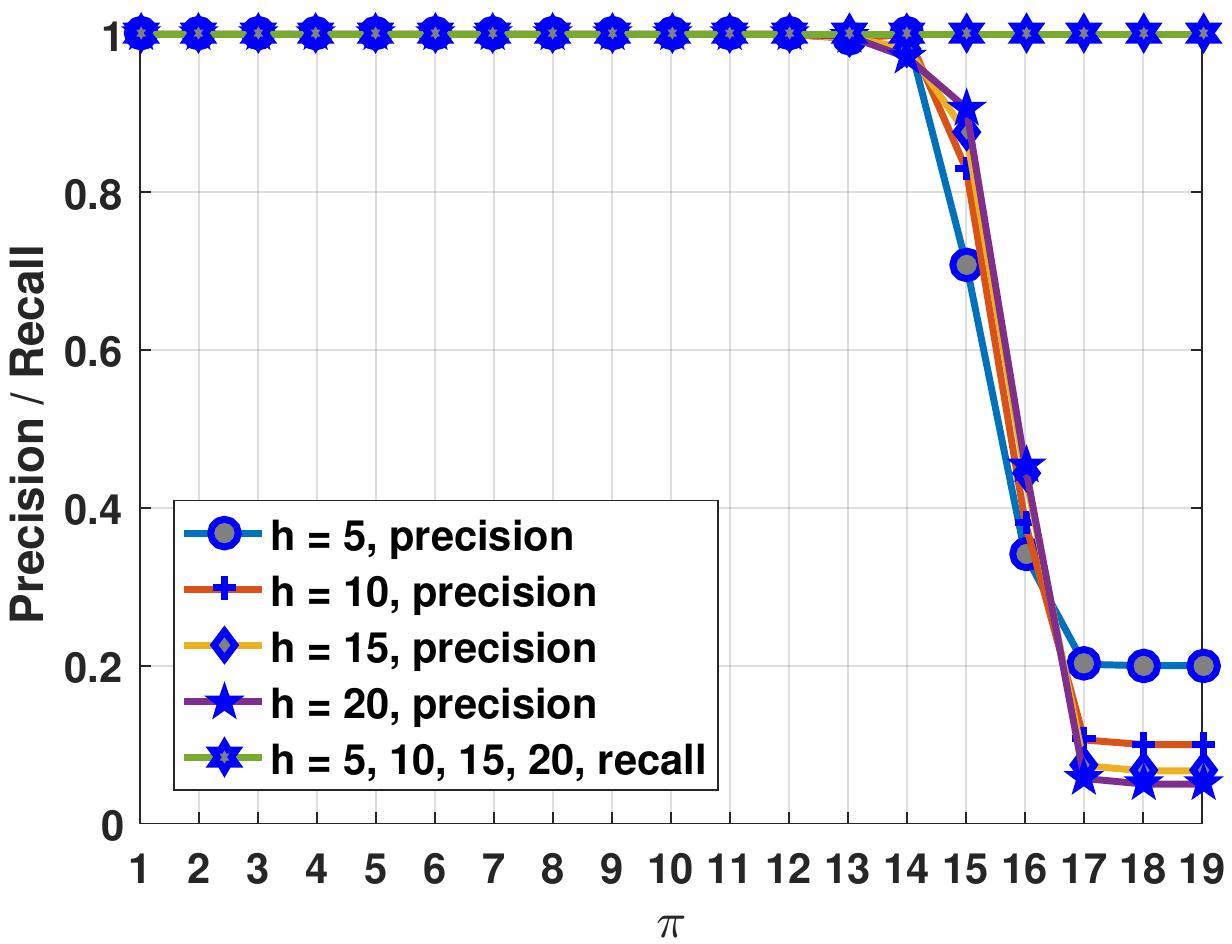}}

	\subfloat[Data owner predicts the number of malicious SPs as $\hat{\phi}$ when $\pi=15$ and the actual number of malicious SPs $\phi=1$. When $h=5$, the maximum value of $\hat{\phi}$ is $5$ as the data owner does not expect more than $h$ malicious SPs.]{\includegraphics[clip,width=0.75\columnwidth]{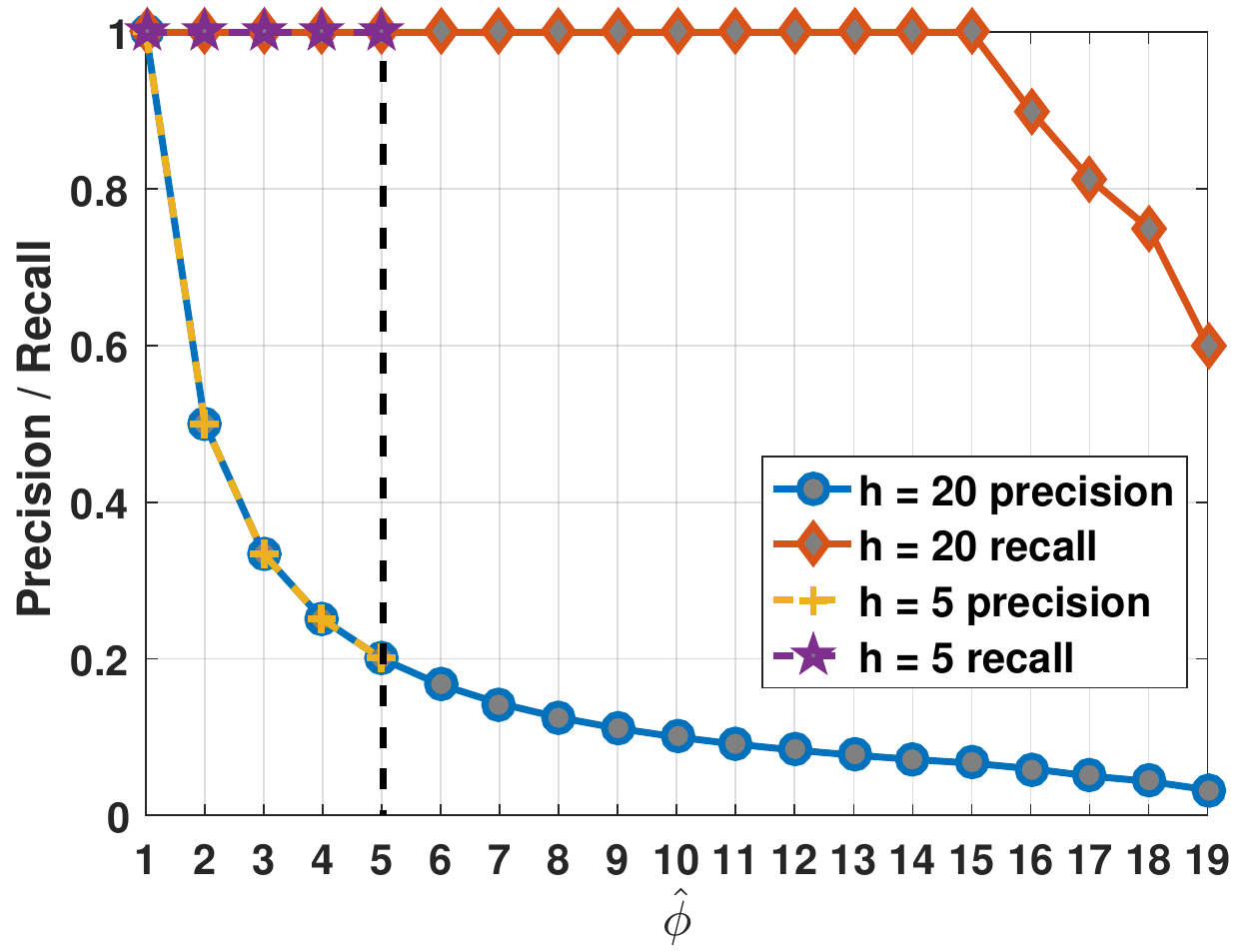}}
	\caption{Precision and recall values for the data owner to detect the malicious SP in the single SP attack in which data has been shared with $h$ SPs. Malicious SP randomly changes ($\pi \times w$) data points to damage the watermark ($w$ is the watermark length).}
	\label{fig:noise_singleSP}
\end{figure}

First, we consider the single SP attack in which data has been shared with $h$ SPs and there is a single malicious SP (data owner Alice may or may not know the number of malicious SPs). Watermark length ($w$) is known by the malicious SP and the malicious SP randomly changes ($\pi \times w$) data points in the data and shares it. For each SP $i$ that received her data, Alice computes $g_i=|Z_\alpha \cap Z_{I_i}|$ ($Z_{I_i}$ is the watermark pattern of SP $i$) and identifies the malicious SP as the one with the highest $g_i$ value. In Figure~\ref{fig:noise_singleSP}(a), we show the precision and recall when the data owner knows that there is a single malicious SP and for different $\pi$ and $h$ values. We conclude that when the data owner knows that there is a single malicious SP, both the precision and recall values are high even when the malicious SP significantly damages the watermark. Under this attack, the proposed scheme is $\rho/\epsilon$-robust against watermark modification with $\rho=\epsilon\simeq1$ when $\pi<13$ and $h\leq20$ ($\pi=13$ means a utility loss of $65\%$).

We also considered the case in which the data owner (Alice) does not know the number of malicious SPs. Here, the actual number of malicious SPs $\phi=1$. We also let $\hat{\phi}$ denote the number of predicted malicious SPs by Alice. This time, Alice predicts the number of malicious SPs as $\hat{\phi}$ ($\hat{\phi}$ can be any number from 1 to $h$). Alice first generates all combinations of $h$ with $\hat{\phi}$. Then, she eliminates the combinations for which the union of the watermarked points of the SPs (in that particular combination) does not contain the watermark pattern in the leaked data ($Z_\alpha$). Next, for each non-eliminated combination $c_i$, she computes $g_i=\sum_{j\in{c_i}}|Z_\alpha \cap Z_{I_j}|$. That is, she computes the sum of intersections of watermarked data points for each SP in the corresponding combination $c_i$ with $Z_\alpha$. Finally, she selects the most likely combination with the highest $g_i$ value and concludes that the SP (or SPs) in the corresponding combination are malicious. In Figure~\ref{fig:noise_singleSP}(b), we set $\pi=15$ and show the precision and recall values for different values of $h$ and $\hat{\phi}$ (when $\phi=1$).\footnote{For smaller values of $\pi$ both precision and recall values are almost equal to $1$.} We observed that when the data owner predicts the actual number of malicious SPs, the actual malicious SP is in set $\mathbf{S}$ with a high probability regardless of the number of sharings of the data owner. Under this attack, the proposed scheme is $\rho/\epsilon$-robust against watermark modification with $\epsilon=1$ when $\pi\leq15$, $h\leq20$, and $\hat{\phi}\leq15$. As expected, precision ($\rho$) decreases as the number of predicted malicious SPs is greater than the actual number of malicious SPs. It is worth noting that this result is obtained when $\pi=15$, which means the malicious SP reduced the utility of the data by $75\%$ to damage the watermark.

\begin{figure}[h!]
\centering
	\subfloat[Data owner knows the number of malicious SPs ($\phi=\hat{\phi}$).]{\includegraphics[clip,width=0.75\columnwidth]{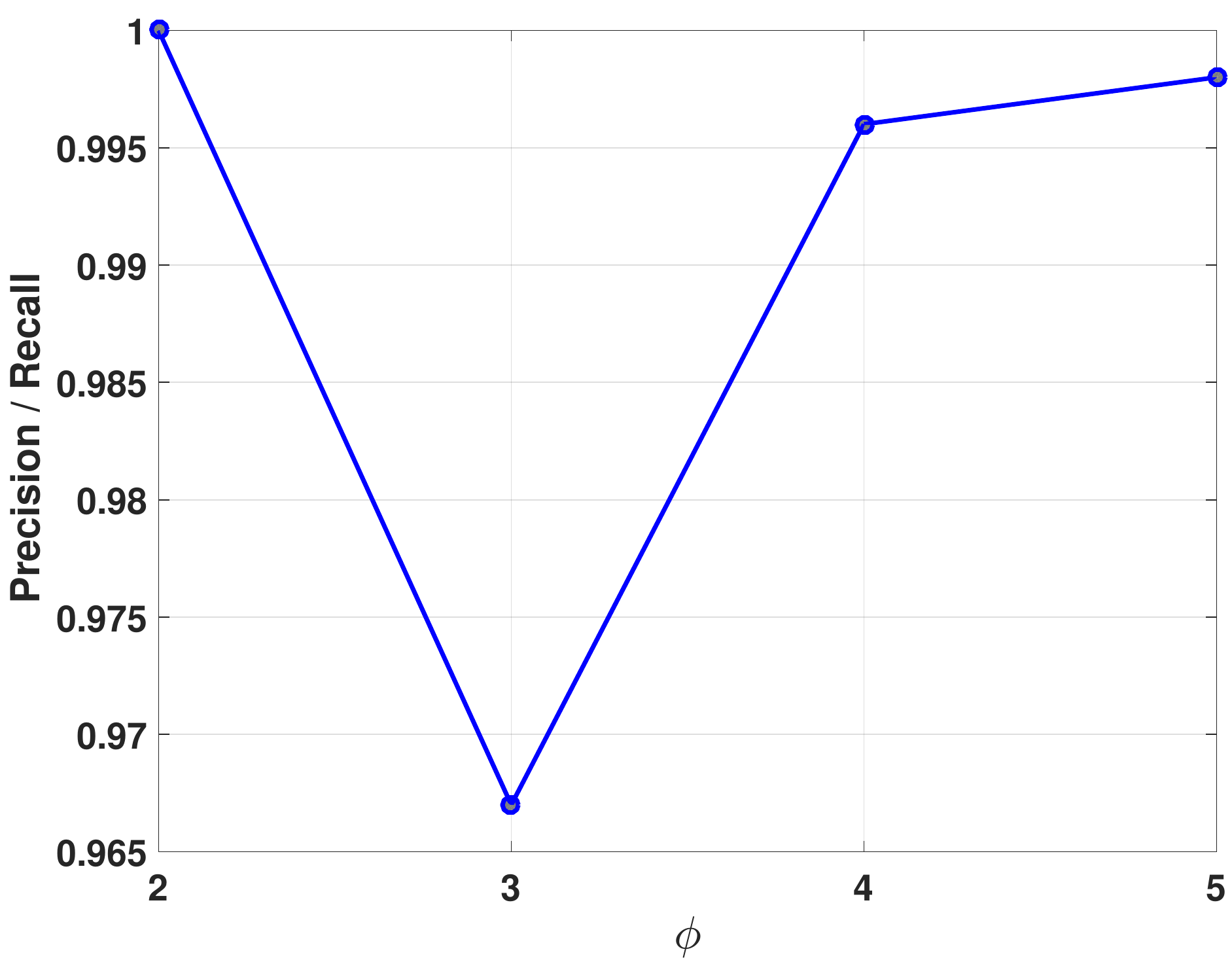}}

	\subfloat[Data owner predicts the number of malicious SPs as $\hat{\phi}$ when the actual number of malicious SPs is $\phi$.]{\includegraphics[clip,width=0.75\columnwidth]{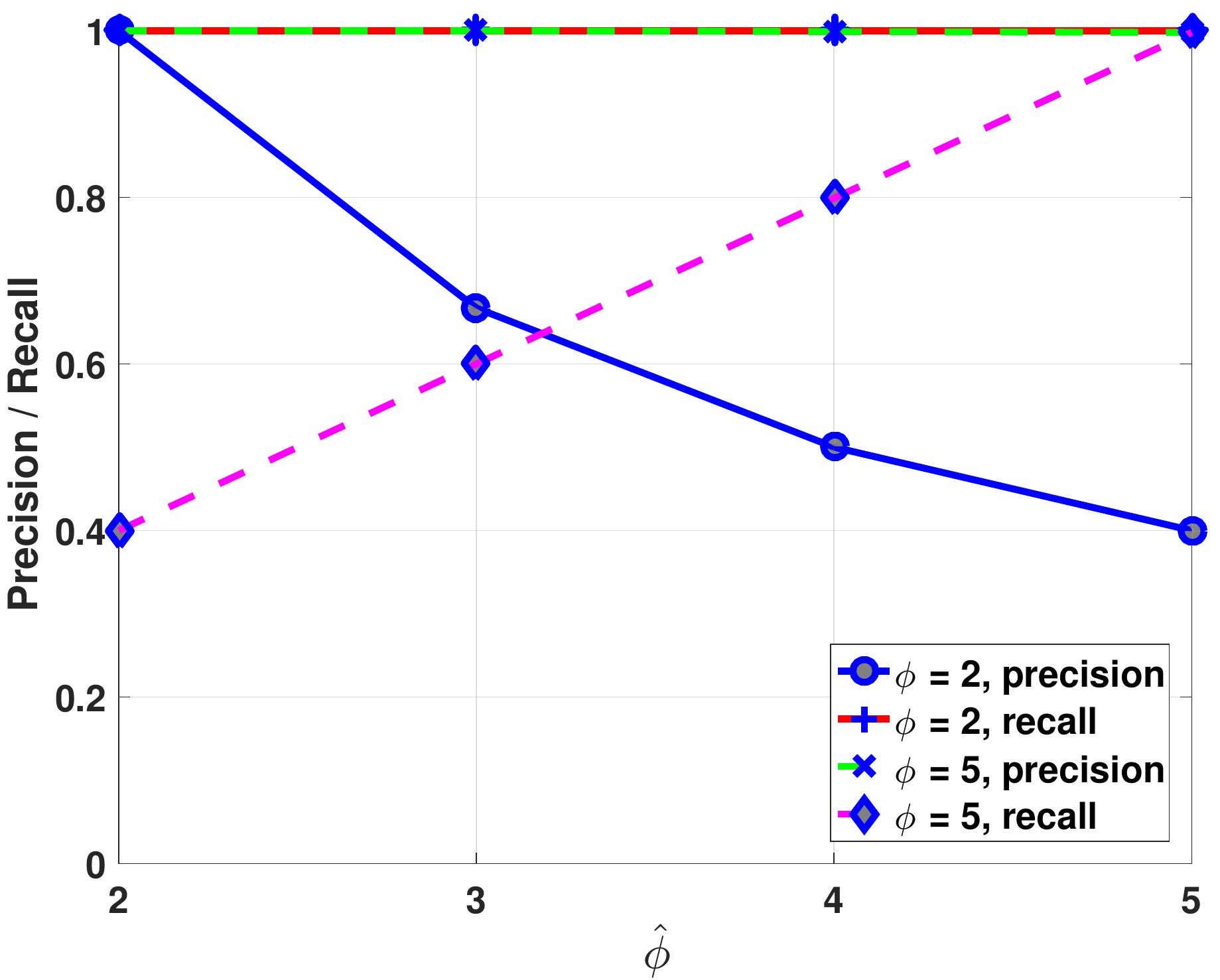}}
	\caption{Precision and recall values for the data owner to detect the malicious SPs in the collusion attack in which data has been shared with $h=10$ SPs. $\phi$ and $\hat{\phi}$ denote the number of actual and predicted malicious SPs, respectively. Malicious SPs only change the states of data points that are different in the aggregated data and do not add further noise ($\pi=0$). In (a), precision and recall curves for different $\phi$ values overlap.}
	\label{fig:noise_collusion}
\end{figure}

Then, we studied the same scenario for the collusion attack. We assume data has been shared with $h=10$ SPs and the number of actual colluding SPs is denoted as $\phi$ ($\phi>1$). Here, malicious SPs compare their aggregated data and only change the states of data points that are different as discussed before. First, we assume that $\phi$ is known by the data owner (i.e., $\hat{\phi}=\phi$). As before, the data owner generates the possible combinations and constructs set $\mathbf{S}$ from the most probable combination. In Figure~\ref{fig:noise_collusion}(a), we show the precision and recall for different $\phi$ values. As before, we also considered the case that the data owner does not know $\phi$ and estimates it as $\hat{\phi}$. In Figure~\ref{fig:noise_collusion}(b), we show the precision and recall for different $\hat{\phi}$ and $\phi$ values. We observed that when the data owner knows the number of actual malicious SPs, we obtain high precision and recall values even when $50\%$ of the SPs that received the data are malicious and colluding. In other words, under this scenario, the proposed scheme is $\rho/\epsilon$-robust against watermark modification with $\epsilon=\rho\simeq0.96$. When the data owner predicts the number of malicious SPs, we observed two cases: (i) when the predicted number of malicious SPs ($\hat{\phi}$) is greater than the actual number of malicious SPs ($\phi$), the proposed algorithm always includes the actual malicious SPs to set $\mathbf{S}$ with a high probability. That is, the proposed scheme is $\rho/\epsilon$-robust against watermark modification with $\epsilon=1$. And, (ii) when $\hat{\phi}$ value is smaller than $\phi$, all the detected malicious SPs in set $\mathbf{S}$ are actual malicious SPs with a very small false positive rate. That is, the proposed scheme is $\rho/\epsilon$-robust against watermark modification with $\rho=1$.

\begin{figure}[h!]
\centering
	\subfloat[Data owner knows the number of malicious SPs ($\phi=\hat{\phi}$).]{\includegraphics[clip,width=0.75\columnwidth]{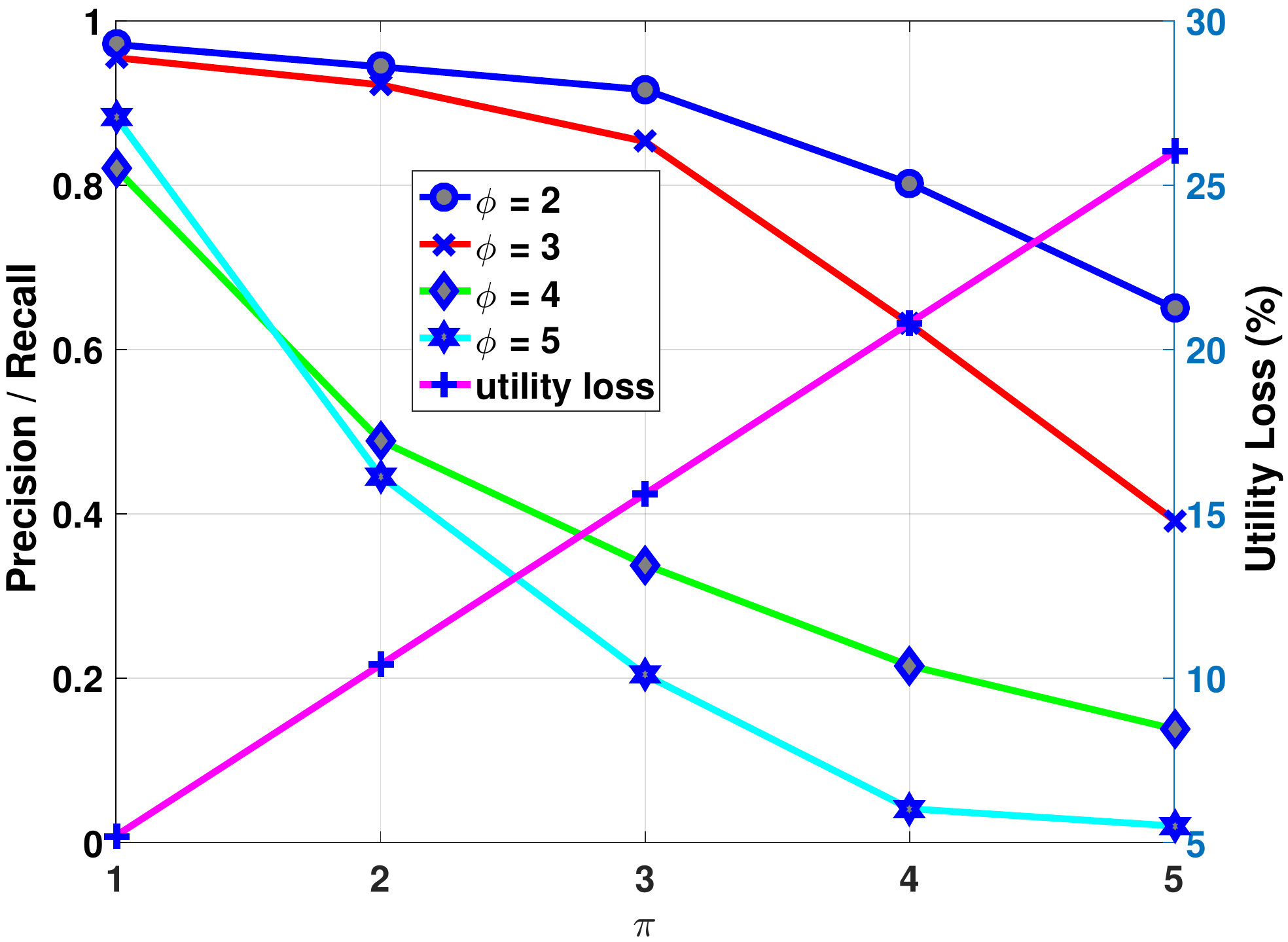}}
	
	\subfloat[Data owner predicts the number of malicious SPs as $\hat{\phi}$ when the actual number of malicious SPs $\phi=5$ and for varying $\pi$ values.]{\includegraphics[clip,width=0.75\columnwidth]{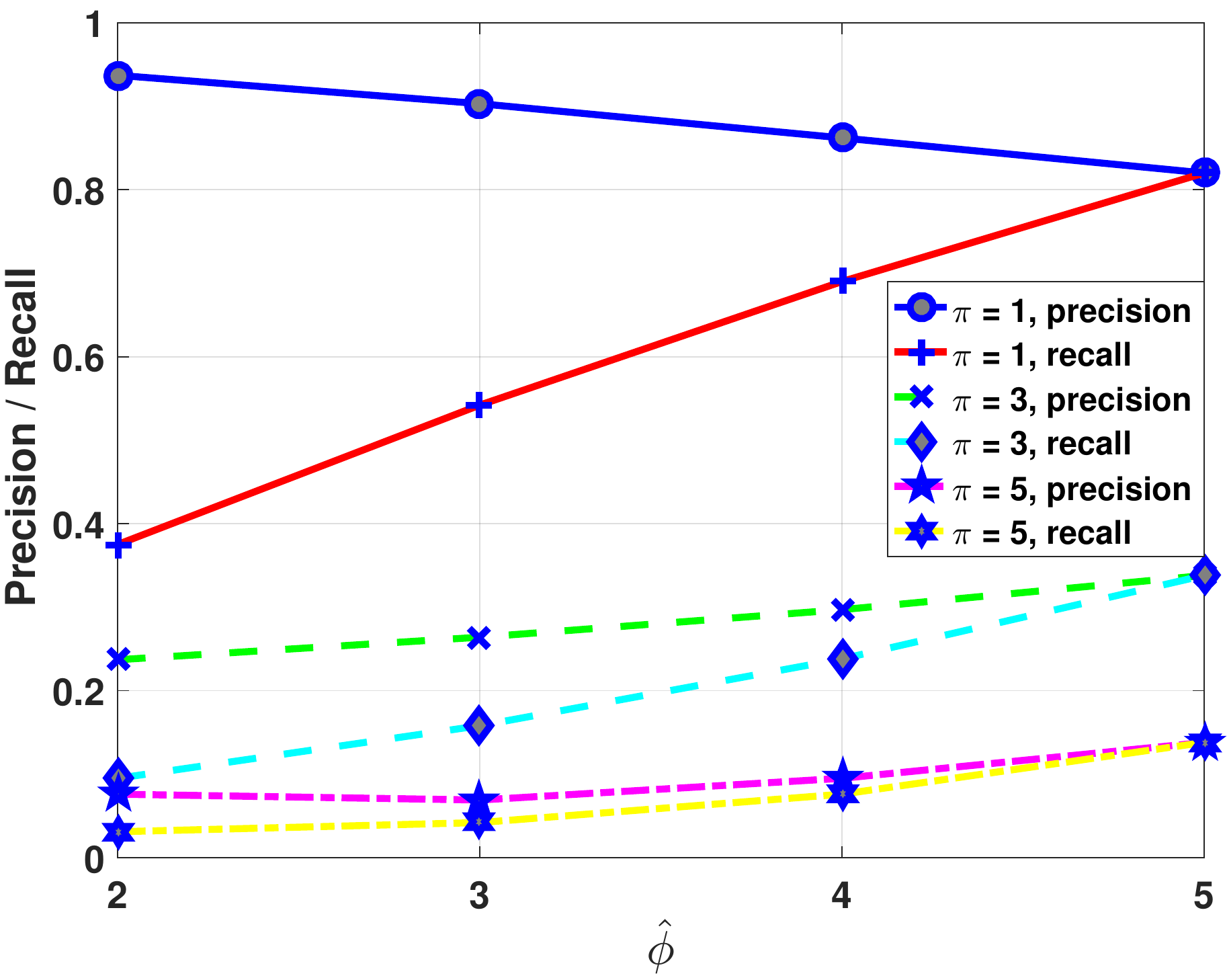}}
	\caption{Precision and recall values for the data owner to detect the malicious SPs in the collusion attack in which data has been shared with $h=10$ SPs. $\phi$ and $\hat{\phi}$ denote the number of actual and predicted malicious SPs, respectively. Malicious SPs both change the states of data points that are different in the aggregated data and they randomly change ($\pi \times w$) data points to damage the watermark ($w$ is the watermark length). In (a), precision and recall curves for different $\phi$ values overlap. Also, in (a), we show the percentage of utility loss due to addition of extra noise by the malicious SPs.}
	\label{fig:noise_collusion_strong}
\end{figure}

Finally, we considered the case in which colluding malicious SPs also add random noise in addition to changing the states of data points that are different in the aggregate data. We assume data has been shared with $h=10$ SPs and colluding malicious SPs randomly change ($\pi \times w$) data points in the data before they leak it. In Figures~\ref{fig:noise_collusion_strong}(a) and~(b), we show the precision and recall when $\hat{\phi}=\phi$, and when the data owner does not know $\phi$, respectively. In Figure~\ref{fig:noise_collusion_strong}(a), we also show the percentage of utility loss in the data due to the noise addition by the malicious SPs (to damage the watermark). Here, the utility loss is shown when $r=0.05$ (i.e., when $5\%$ of original data is watermarked). As $r$ value increases, the loss in utility (due to extra noise addition by the malicious SPs) also increases linearly. For instance when $r=0.1$, to decrease the precision and recall values down to $0.2$, half of the SPs that received the data should be malicious and they need to add noise to $50\%$ of the original data to damage the watermark. As shown in Figure~\ref{fig:noise_collusion_strong}(a), if the data owner knows the number of malicious SPs, both precision and recall of detection performance are high up to $30\%$ of the SPs that received the data are malicious (and colluding) and up to a utility loss of $15\%$. That is, the proposed scheme is $\rho/\epsilon$-robust against watermark modification with $\epsilon=\rho\simeq0.9$ up to $\phi=3$ and $\pi=3$. Beyond this, we observed a decrease in both precision and recall with increasing $\pi$ and $\phi$ values. This behavior gives some idea about the practical limits of our proposed scheme. When data owner predicts the number of malicious SPs (Figure~\ref{fig:noise_collusion_strong}(b)), we observed two cases: (i) when the added noise by the malicious SPs is less than $3$ times the watermark length, the proposed scheme includes the actual malicious SPs in set $\mathbf{S}$ with a high probability. That is, the proposed scheme is $\rho/\epsilon$-robust against watermark modification with $\rho\simeq0.7$ up to $\pi=3$ and for all $\hat{\phi}$ values. When the added noise by malicious SPs is beyond this value, both precision and recall values start decreasing. Note however that adding noise beyond this value significantly reduces data utility as discussed before.

\color{black}

\section{Discussion}\label{sec:discussion}

\color{black}

Here, we discuss potential use of our proposed scheme in real-life, its potential extensions, and future research directions.

\noindent{\textbf{Usability.}} The proposed system detects the malicious SPs if data is leaked or sold without the data owner's consent and if the data owner observes this leakage. Similarly, the SP that buys the data may keep the malicious SPs liable from this unauthorized sharing (with the cooperation of the data owner). It may be practically infeasible for a data owner to notice her data is leaked at the first place. Instead, this can be outsourced to a third party that continuously analyzes publicly available datasets that are made available by SPs that collect personal information.

\noindent{\textbf{Attacker's auxiliary information.}} Security of the proposed watermarking scheme also depends on the auxiliary knowledge of the malicious SPs. For example, genomes of family members are highly correlated with each other. Therefore, if a malicious SP obtains genomic data from multiple individuals from the same family, it may have higher probability to determine the watermarked points by analyzing the expected correlations across such individuals' data. In future work, we will extend our optimization problem by also considering such correlations across different data owners. Similarly, if the malicious SP knows phenotypic information about the data owner, it may connect this information to her genomic data. In general, auxiliary information about the data owner may help a malicious SP to infer the watermarked positions with higher probability. We will also study the impact of background information a malicious SP may have about an individual to the security of the proposed watermarking scheme.

\noindent{\textbf{Data privacy.}} Many recent developments in data sharing services are based on users adding noise to the data they send to the SPs (e.g., location data is perturbed before it is shared with a location-based service provider~\cite{Andres:2013:GDP:2508859.2516735}). By integrating the differential privacy concept~\cite{Dwork:2006:DP:2097282.2097284} to our proposed watermarking scheme, we can actually both add watermark and use the watermark as noise to also provide privacy. Traditionally, differential privacy concept is applied to statistical databases. Recently, differential privacy concept has been used for individual release of location data~\cite{Andres:2013:GDP:2508859.2516735}. As future work,  we will integrate the idea in~\cite{Andres:2013:GDP:2508859.2516735} into our watermarking algorithm.

In this work, we assume that data is shared with the SPs in a non-anonymous way. On the other hand, if data is shared by the data owner in an anonymous way (e.g., by removing the identity and applying perturbation), the SPs first need to find the connection between data portions they received from the same data owner. This will create another challenge for the malicious SPs to perform the collusion attack and we will analyze this scenario in the future work.

\noindent{\textbf{Data utility.}} We show that in all attacks, utility remains high (i.e., number of watermarked positions are significantly less). We obtained this by trying different $r$ values (i.e., ratio of the watermark length to the data length). We can alternatively include $w$ (i.e., watermark length) as one of the objectives of the optimization problem and put a limit on it. When we include $w$ in the objective function, the problem becomes a multi-objective optimization problem. Solution of a multi-objective optimization problem is non-trivial and many proposed techniques suggest converting the multi-objective problem into a single-objective one. Thus, we transform this multi-objective problem into single objective problem.

In this new formulation, there are two additions to the optimization problem introduced in Section~\ref{sec:For Data With No Correlations}. First, the objective function is changed as follows:
\begin{equation}
	min \{\beta \cdot \prod_{i=0}^{h+1} (\frac{n_i^{h+1}}{n_i^{h+1} + n_{h-i+1}^{h+1}})^{n_i^{h+1}} + (1-\beta) \cdot w\}\nonumber
\end{equation}
We use the weighted sum of the watermark length and the inference probability as the new objective function. The weight ($\beta$) determines the tradeoff between the inference probability and the watermark length (i.e., data utility). Second, we keep all 7 constrains the same (as in Section~\ref{sec:For Data With No Correlations}) and add a new constraint as $w < w_m$, where $w_m$ is the maximum allowed watermark length. This new constraint puts a threshold to the maximum number of watermark points. This new optimization problem guarantees the minimum weighted sum of inference probability and watermark length.

Depending on the data type, other utility constraints may also be included in the proposed algorithm. For instance, if adding watermark to two consecutive data points significantly reduces data utility, once $y_i^h$ and $\hat{y}_i^h$ values are determined as a result of the optimization problem, watermark addition algorithm in Section~\ref{sec:For Data With No Correlations} (or Section~\ref{sec:For Data With Correlations}) can be tailored to take this constraint into account while adding the watermarks.


\color{black}

\noindent{\textbf{Data credibility.}} Our scheme guarantees that if data is leaked, the owner of the data will be able to find the source of the leakage. However, SPs that receive data from an individual should also be sure that the data is credible. Up to this point, we do not guarantee the credibility of the data. A malicious individual may mislead an SP by sending fake data to it. Ayday et al. proposed a cryptographic system that guarantees data credibility while sharing genomic data between entities~\cite{Ayday2017}. The system in~\cite{Ayday2017} uses homomorphic signatures and aggregate signatures and it consists of three parties: (i) data owner (Alice), (ii) the certified institution (CI), and (iii) the service provider (SP). The role of the data owner and the SP is the same with our proposed system. We can apply a simplified version of the scheme in~\cite{Ayday2017} to also provide data credibility with the help of a CI. In that scenario, data owner (Alice) sends her data points to the CI and the CI signs each data point individually by using a homomorphic signature scheme. Alice, when she shares her data with an SP, computes a signature on a linear combination of her shared data using the homomorphic properties of the signature scheme. Thus, given the data points it receives from Alice and the public key of the CI, the SP can verify the credibility of data by verifying the signature.

As future work, we will integrate this solution to our proposed watermarking algorithm to address the data credibility issue. With this technique, a malicious SP cannot partially share Alice's data or malicious SPs cannot modify Alice's data to damage the watermark since data credibility will be lost in both cases. Therefore, attacks presented in Section~\ref{sec:detection_performance} would not be possible as the malicious SPs cannot prove the credibility of the data once they modify it.

\color{black}

\noindent{\textbf{Other applications.}} The proposed watermarking algorithm can be applied for any type of sequential data (we describe the general framework for sequential data in Section~\ref{sec:Proposed Solution}). However, implementation for different data types is non-trivial. Watermarked data point is not always changed to a predetermined state; there may be many alternatives (e.g., perturbed location may have many different states that are determined based on privacy and utility). Furthermore, correlations in other types of data may be more complex. To address some of these challenges, we will work on the application of the proposed scheme for location patterns as future work.

\color{black}

\section{Conclusion} \label{sec:Conclusion}
In this work, we have proposed a scheme to share sequential data while addressing the liability issues in case of unauthorized sharing. The proposed scheme is between a data owner and one or more service providers. We have shown that the proposed watermarking scheme provides high security against collusion and correlation attacks. That is, with high probability, malicious service providers cannot identify the watermark on the data even if they collude or try to use the inherent correlations in the data. We have also shown that the proposed scheme does not degrade the utility of data while it provides the aforementioned security guarantees. We believe that the proposed work will deter the service providers from unauthorized sharing of personal data with third parties.

\end{document}